\documentclass[fleqn,usenatbib]{mnras}

\usepackage{newtxtext,newtxmath}

\usepackage[T1]{fontenc}

\usepackage{graphicx}	
\usepackage{amsmath}	
\usepackage{siunitx}    
\usepackage{bm}
\usepackage{multirow}
\usepackage{xcolor}

\newcommand{\logNHI}{\ensuremath{\log N(\mbox{\ion{H}{i}})}}
\newcommand{\logNHIunit}{\ensuremath{\log [N(\mbox{\ion{H}{i}})/\rm{cm}^{-2}]}}
\newcommand{\kms}{\ensuremath{\text{km s}^{-1}}\,}
\newcommand{\Moyr}{\ensuremath{\text{M}_{\odot}\ \text{yr}^{-1}}}
\DeclareUnicodeCharacter{0308}{}
\newcommand\bolden[1]{{\boldmath\bfseries#1}}

\title[Gas Flows in the CGM]{MUSE-ALMA Halos XI: Gas flows in the circumgalactic medium}

\author[Weng et al.]{
Simon Weng,$^{1,2,3,4}$\thanks{E-mail: simonw358@gmail.com}
{C\'eline P\'eroux},$^{1,5}$
Arjun Karki,$^{6}$ 
Ramona Augustin,$^{7}$ 
Varsha P. Kulkarni,$^{6}$ 
\newauthor
Aleksandra Hamanowicz,$^{7}$
Martin Zwaan,$^{1}$ 
Elaine M. Sadler,$^{2, 3, 4}$
Dylan Nelson,$^{8}$ 
Matthew J. Hayes,$^{9}$ 
\newauthor
Glenn G. Kacprzak,$^{10, 3}$
Andrew J. Fox,$^{11, 12}$ 
Victoria Bollo,$^{1}$
Benedetta Casavecchia,$^{13}$
Roland Szakacs,$^{1}$
\\
$^{1}$ European Southern Observatory, Karl-Schwarzschildstrasse 2, D-85748 Garching bei M{\"u}nchen, Germany\\
$^2$ Sydney Institute for Astronomy, School of Physics A28, University of Sydney, NSW 2006, Australia\\
$^3$ ARC Centre of Excellence for All Sky Astrophysics in 3 Dimensions (ASTRO 3D)\\ 
$^4$ ATNF, CSIRO Space and Astronomy,  PO Box 76, Epping, NSW 1710, Australia \\
$^{5}$ Aix Marseille Universit\'e, CNRS, LAM (Laboratoire d'Astrophysique de Marseille) UMR 7326, 13388, Marseille, France \\
$^6$ Department of Physics and Astronomy, University of South Carolina, Columbia, SC 29208, USA\\
$^{7}$ Space Telescope Science Institute, 3700 San Martin Drive, Baltimore, MD 21218, USA \\
$^{8}$ Universit\"at Heidelberg, Zentrum f\"ür Astronomie, Institut f\"ur theoretische Astrophysik, Albert-Ueberle-Str. 2, 69120 Heidelberg, Germany\\
$^{9}$ Stockholm University, Department of Astronomy and Oskar Klein Centre for Cosmoparticle Physics, AlbaNova University Centre, SE-10691, Stockholm, Sweden\\
$^{10}$ Centre for Astrophysics and Supercomputing, Swinburne University of Technology, Hawthorn, Victoria 3122, Australia\\
$^{11}$ AURA for ESA, Space Telescope Science Institute, 3700 San Martin Drive, Baltimore, MD 21218, USA \\
$^{12}$ Department of Physics \& Astronomy, Johns Hopkins University, 3400 N. Charles St., Baltimore, MD 21218 \\
$^{13}$ Max-Planck-Institut f\"{u}r Astrophysik, Karl-Schwarzschild-Strasse 1, D-85748 Garching b. M\"{u}nchen, Germany
}

\date{Accepted XXX. Received YYY; in original form ZZZ}

\pubyear{2022}

\begin{document}
\label{firstpage}
\pagerange{\pageref{firstpage}--\pageref{lastpage}}
\maketitle

\begin{abstract}
The flow of gas into and out of galaxies leaves traces in the circumgalactic medium which can then be studied using absorption lines towards background quasars. 
We analyse 27 $\logNHIunit > 18.0$ \ion{H}{i} absorbers at $z = 0.2$ to $1.4$ from the MUSE-ALMA Halos survey with at least one galaxy counterpart within a line of sight velocity of $\pm 500$ \kms\!. 
We perform 3D kinematic forward modelling of these associated galaxies to examine the flow of dense, neutral gas in the circumgalactic medium. 
From the VLT/MUSE, HST broadband imaging and VLT/UVES and Keck/HIRES high-resolution UV quasar spectroscopy observations, we compare the impact parameters, star-formation rates and stellar masses of the associated galaxies with the absorber properties. 
We find marginal evidence for a bimodal distribution in azimuthal angles for strong \ion{H}{i} absorbers, similar to previous studies of the \ion{Mg}{ii} and \ion{O}{vi} absorption lines. 
There is no clear metallicity dependence on azimuthal angle and we suggest a larger sample of absorbers are required to fully test the relationship predicted by cosmological hydrodynamical simulations. 
A case-by-case study of the absorbers reveals that ten per cent of absorbers are consistent with gas accretion, up to 30 per cent trace outflows while the remainder trace gas in the galaxy disk, the intragroup medium and low-mass galaxies below the MUSE detection limit. 
Our results highlight that the baryon cycle directly affects the dense neutral gas required for star-formation and plays a critical role in galaxy evolution. 
\end{abstract}

\begin{keywords}
galaxies: evolution -- galaxies: formation -- galaxies: kinematics and dynamics -- galaxies: haloes -- quasars: absorption lines
\end{keywords}



\section{Introduction}
Once known as `island universes', galaxies are no longer considered isolated systems, especially when observing the cycle of baryons within these systems \citep{PerouxHowk2020}. 
The formation of stars is mediated by the delicate balance between outflowing and inflowing gas. 
Galactic winds driven by active galactic nuclei (AGN) or intense star-formation remove gas from galaxies \citep{Veilleux2005}. 
Some of the outflowing material condenses back onto the galaxy in the form of fountains \citep{Fraternali2008, Marinacci2010, Fraternali2017}. 
At the same time, the accretion of cold gas from large-scale filaments and the cooling of hot halo gas replenishes the gas reservoirs of galaxies \citep{Keres2005, Nelson2013, Hafen2022}. 
Satellites embedded within the halo of galaxies are also an important source of cool gas \citep{Wang1993, Hafen2019}. 
These phenomena leave their traces in the circumgalactic medium (CGM), the region that extends from the galactic disk to the intergalactic medium \citep[IGM;][]{Tumlinson2017, Faucher2023}.

While the diffuse gas in the CGM can be studied in emission, it is typically detected in systems containing quasars, extreme starbursts and overdensities not representative of typical galaxies \citep[e.g.][]{Hayes2016, Epinat2018, Johnson2018, Chen2019, Helton2021, Burchett2021, Cameron2021}. 
Stacking enables detections of the diffuse gas in emission, but it becomes difficult to link the detections to galaxy properties and the wider environment \citep{Steidel2011, Momose2014, Chen2020b, Dutta2023}. 
Typical galaxies on the main sequence require exceptional exposure times to obtain a detection in emission \citep{Zabl2021, Leclercq2022, Bacon2023}. 
Alternatively, analysing absorption lines in the spectra of bright background sources such as quasi-stellar objects (QSOs) allows us to probe the gas in the CGM to lower column densities. 
While these sightlines only sample the CGM on parsec scales, there has been an abundance of recent surveys examining the galaxies around \ion{H}{i} \citep[][Karki et al. 2023]{Lofthouse2020, Muzahid2020, Chen2020, Berg2023}, \ion{Mg}{ii} \citep{Bouche2016, Nielsen2020, Lundgren2021} and other gas phases. 

Early studies of strong \ion{H}{i} Ly-$\alpha$ absorbers associated them to the rotating disks of galaxies \citep{Wolfe1986, Wolfe2005}. 
Damped Ly-$\alpha$ absorbers (DLAs) with column density $\logNHIunit \geq 20.3$ are expected to originate from gas within $\sim$20 kpc of the galaxy centre \citep{Peroux2005, Zwaan2005, Stern2021}. 
Since, the increasing number of galaxy-absorber systems brought by the advent of integral field spectroscopy suggest these absorbers can originate from beyond the disk. 
In particular, \ion{H}{i} absorbers with lower column density found at larger impact parameters suggest different phenomena such as gas accretion and outflows are being traced. 

The time galaxies take to deplete their molecular gas reservoirs is found to be shorter than the Hubble time at all redshifts, meaning that galaxies must have a way to maintain a gas supply through accretion \citep{Daddi2010, Genzel2010, PerouxHowk2020, Tacconi2020, Walter2020}. 
Some theoretical studies have attempted to separate gas accretion into cold-mode and hot-mode accretion. 
The former refers to the accretion of cold gas from filaments in the intergalactic medium and is expected to dominate at redshift $z \gtrsim 3$ \citep{Keres2005, Nelson2013}. 
Simulations predict that the accretion produces flows that co-rotate with the galaxy disk out to $\sim$100 kpc \citep{Stewart2011, Stewart2013, vandeVoort2011} and can be traced by dense \ion{H}{i} absorbers \citep{vandevoort2012, Theuns2021}. 
In contrast, the cooling of hot halo gas is also expected to grow gas reservoirs and dominates at lower redshifts and/or higher masses \citep{Dekel2009a, Dekel2009b}. 
The accretion of hot gas transitions from a spherical geometry to a disk when it cools because of angular momentum conservation \citep{Mo1998, Hafen2022}. 
While gas accretion at $z \lesssim 1$ has been probed using \ion{Mg}{ii} absorbers \citep{Ho2017, Martin2019, Zabl2019}, studies of accretion using Ly-$\alpha$ absorbers remain limited. 

In contrast to accretion, neutral gas in outflows has been ubiquitously observed in galaxies when the SFR surface density is sufficiently high \citep[e.g.,][]{Heckman2017, Hayes2023} and in galaxies containing an active galactic nucleus \citep[e.g.,][]{Cicone2014, Cicone2015}.
While there remains uncertainty regarding how the cool gas survives in the hot and turbulent wind environment, recent idealised simulations of the cold phase in a turbulent medium find that cold gas clouds can survive and even grow depending on their size \citep{Gronke2022}. 
Traditionally, outflowing gas has been observed using down-the-barrel observations by identifying blueshifted absorption against the galaxy stellar continuum \citep{Martin2005, Veilleux2005, Rubin2014, Heckman2017}. 
Identifying outflows using absorbers towards background sources is a complementary method where the location of the absorbing gas is precisely measured \citep{Bouche2012, Kacprzak2014, Schroetter2019}. 
However, the velocity degeneracy between inflows and outflows increases the difficulty in identifying gas flows for these transverse absorption-line studies. 
Observations and simulations of gas outflows that emerge from the galaxy centre suggest they form an expanding biconical shape perpendicular to the galaxy disk as this is the path of least resistance \citep{Bordoloi2011, Lan2014, Nelson2019b}. 
Hence, quasar absorbers have been interpreted to be outflowing when they are aligned with the projected minor axis of galaxies \citep{Rahmani2018b, Schroetter2019}. 
Transverse absorption-line studies of outflowing gas at $z > 0.2$ typically use the \ion{Mg}{ii} ion, but there have been limited studies on Ly-$\alpha$ absorbers that directly trace the neutral hydrogen required for star-formation. 

On the other hand, Ly-$\alpha$ absorbers at $z \lesssim 1$ have a bimodal distribution in the gas-phase metallicity \citep{Lehner2013, Lehner2022a}. 
Here, metal-poor systems are expected to trace the accretion of cold gas, while metal-rich absorbers trace outflows, recycled or tidally stripped gas \citep{Peroux2020, Lehner2022a}. 
Indeed, studies of high velocity clouds (HVCs) in the Milky Way supplement velocity measurements of the HVC with metallicity to distinguish between inflows and outflows \citep{Fox2019, Ramesh2023}. 
For extragalactic sources, the imprint of inflows and outflows in the CGM might also be seen in the bimodal distribution of azimuthal angles ($\Phi$) between a galaxy's major axis and the absorber sightline \citep[where values close to $90^\circ$ indicate the gas is located near the projected galaxy minor axis;][]{Bouche2012, Kacprzak2012, Bordoloi2014, Schroetter2019}. 
Naturally, it follows that there should be a metallicity dependence on azimuthal angle, where gas located towards the minor axis is metal-enriched as outflows eject metals produced in stars \citep{Nelson2019b, Peroux2020, Truong2021, vandevoort2021}. 
However, in observations, this phenomenon remains unclear as studies have found no such trend \citep{Peroux2016, Kacprzak2019, Pointon2019}, perhaps due to uncertainties in the ionization correction and dust modelling of absorbers or poor metal-mixing in the CGM \citep{Zahedy2019, Zahedy2021, Bordoloi2022, Sameer2022}. 
Simulations of \ion{Mg}{ii} absorption around galaxies find that the equivalent width (EW) of the line decreases as the azimuthal angle of the sightline increases for a fixed impact parameter \citep{DeFelippis2021}. 
Interestingly, \citet{Wendt2021} finds gas located near the minor axis is more dust-depleted and perhaps metal-enriched than absorbers near the major axis. 
More carefully selected observational samples are required to reach a consensus on the metallicity--azimuthal angle dependence, especially because the relationship is dependent on the galaxy stellar mass, impact parameter, redshift and \ion{H}{i} column density \citep{Peroux2020}. 

The MUSE-ALMA Halos survey targets 32 Ly-$\alpha$ absorbers with column densities $\logNHIunit > 18.0$ at redshift $0.2 \lesssim z \lesssim 1.4$ \citep[see][for an overview]{Peroux2022}. 
In total, 79 galaxies are detected within $\pm 500$ \kms of the absorbers at impact parameters ranging from 5 to 250 kpc \citep{Weng2022}. 
The stellar masses of these associated galaxies have been measured from broadband imaging in Augustin et al. (in prep) and their morphologies studied in Karki et al. (submitted). 
Gas flows for a subsample of galaxies have already been studied in earlier works \citep{Peroux2017, Klitsch2018, Rahmani2018a, Rahmani2018b, Hamanowicz2020, Szakacs2021} and our work is a continuation of these studies for the full MUSE-ALMA Halos survey. 
With the complete sample of absorbers and their galaxy counterparts, we examine the azimuthal dependence of metallicity in the circumgalactic medium and identify the gas flows being probed by QSO sightlines. 
We adopt the following $\Lambda$CDM cosmology: $\rm H_0 = 70$ \kms $\rm {Mpc}^{-1}$, $\Omega_M = 0.3$ and $\Omega_\Lambda = 0.7$.

\section{Ionized Gas Maps}
\label{sec: galpak}
To determine the origin of the gas, that is, whether it traces inflows, outflows or other phenomena, we need to compare the ionized gas kinematics with the neutral-phase absorber kinematics. 
We create continuum-subtracted cubes centred on the [\ion{O}{ii}]$\lambdaup \lambdaup 3727,3729$, H$\beta$ $\lambdaup 4861$, [\ion{O}{iii}] $\lambdaup 5007$ and H$\alpha$ $\lambdaup 6563$ emission lines for each galaxy when available within the MUSE wavelength range ($4700$ to $9300$ \AA). 
We use the \textsc{galpak} algorithm v1.32 to extract intrinsic galaxy parameters such as the inclination ($i$) and kinematic position angle (PA$_{\rm kin}$) and velocity maps from data cubes. 
This forward modelling approach assumes a disk model and we refer readers to \citet{Bouche2015} for a complete description of the method. 
For the code to provide stable results, the signal-to-noise ratio (S/N) of the nebular line at the brightest spaxel needs to meet the requirement S/N $> 3$.
Additionally, the algorithm is robust provided the galaxy is not too compact; the prerequisite $R_{e}/R_{\rm PSF} > 1.50$ should be met, where $R_{\rm PSF}$ is half the full width at half maximum (FWHM) of the PSF and $R_{e}$ is the effective radius.

From the initial sample of 79 associated galaxies, 48 satisfy the signal-to-noise ratio and projected galaxy size criteria. 
We run the algorithm for all the nebular lines ([\ion{O}{ii}], H$\beta$, [\ion{O}{iii}] and H$\alpha$). 
For galaxies where multiple emission lines are available, we find the measured parameters agree within 10 per cent for any given galaxy. 
We preferentially adopt the values returned by the H$\beta$, [\ion{O}{iii}] and H$\alpha$ lines to avoid complications associated with fitting the [\ion{O}{ii}] doublet. 
The maximum rotational velocity, inclination and kinematic position angle are tabulated in \autoref{tab:kin_properties}.
We also include photometric position angles (PA$_{\rm phot}$) from \textsc{galfit} fits of the galaxy using the HST broadband imaging in the online version of the table (Karki et al. submitted). 

The smoothed rotational velocity maps for three example galaxies are shown in \autoref{fig:Maps1} and the remainder are found in Appendix \ref{app:maps} (Figures \ref{fig:Maps2} to \ref{fig:Maps10}). 
Each map is accompanied by dashed lines marking projected angles of $\pm 30^{\circ}$ away from the minor axis. 
The arrow points in the direction of the QSO sightline and consists of two halves coloured by the impact parameter and line of sight velocity difference between galaxy and absorber. 
All galaxies located at the smallest impact parameter to the absorber have their velocity maps emboldened. 
We provide the emission line used to fit each galaxy, identification and absorber redshift below the velocity map. 
For each fitted galaxy, we also display the observed flux and flux residual maps from the \textsc{galpak} fit. 
Regions of the flux map that have a white background are pixels that were masked during fitting or are outside the MUSE field of view. 
Pixels were masked to remove the flux contribution from the nearby QSO. 
The arrow below the residual maps demarcates the physical size of the cube used to fit the emission. 

Ten galaxy fits have significant flux residuals and can be attributed to several causes. 
The residuals for four galaxies in the Q1130$-$1449 field associated with the absorber at $z_{\rm abs} = 0.3127$ arise because of an extended ionized gas nebula permeating the large galaxy group \citep{Kacprzak2010, Chen2019, Peroux2019}. 
Galaxies Q1130$-$1449\_6 and Q1130$-$1449\_8 additionally have significant residuals due to a possible merger and outflows respectively.   
There are also significant flux residuals for galaxy Q0454$-$220\_4 due to the flux saturation of IFUs in the MUSE data from a nearby bright star. 
Another two galaxies (Q1211$+$1030\_7 and Q1229$-$021\_6) likely host active galactic nuclei from their positions on the [\ion{O}{iii}] $\lambdaup 5007$/H$\beta$ versus [\ion{O}{ii}]$\lambdaup \lambdaup 3727,29$/H$\beta$ classification diagram \citep{Lamareille2010, Weng2022}. 
As \textsc{galpak} compares a disk model to the data using forward modelling, the flux excess is likely caused by the AGN. 
Finally, the significant residuals at the centre of source Q0454$-$220\_69 are a result of flux contamination from the QSO less than two arcseconds away. 
We exclude these objects with unreliable fits from further analysis and but include and note them in the figures for completeness. 


\begin{figure*}
    \includegraphics[width=\textwidth]{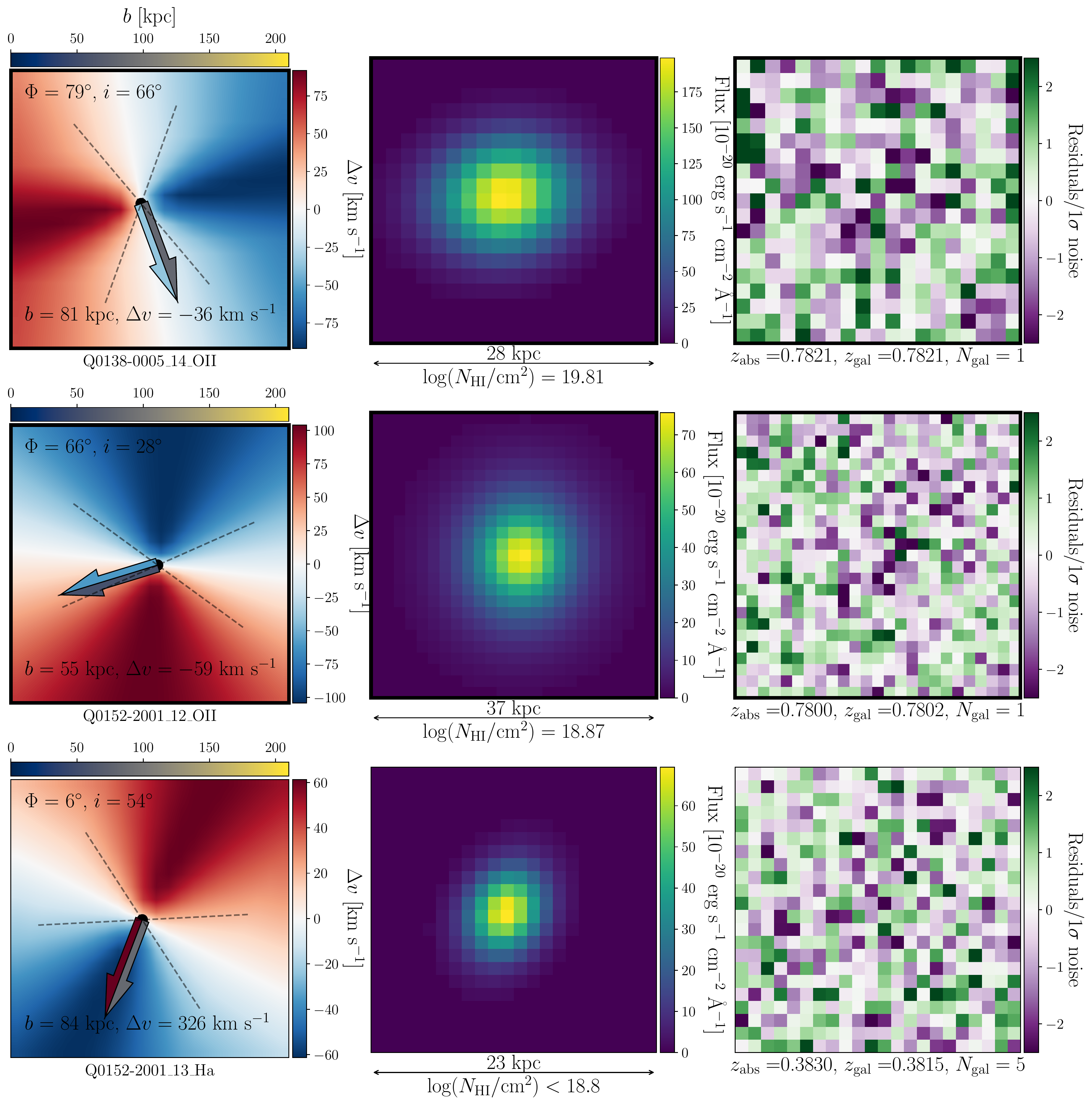}
    \caption{The modelled velocity, observed flux and residual maps of three randomly chosen galaxies associated with QSO absorbers that were fitted using \textsc{galpak}. 
    Each galaxy is represented by three maps. 
    The modelled velocity maps in column one have been smoothed using repeated linear interpolation. 
    A black dot indicates the modelled galaxy centre and two dashed lines represent 2D projected angles of $\pm 30^\circ$ from the minor axis. 
    The arrow points in the direction of the \ion{H}{i} absorber and consists of two coloured halves. 
    The halves are coloured by the impact parameter and line of sight velocity difference between galaxy and absorber. 
    When pointing up, the left half corresponds to the impact parameter and the reverse is true when pointing down. 
    We include text with the azimuthal angle ($\Phi$) and inclination ($i$) along with the impact parameter ($b$) and line of sight velocity difference ($\Delta v$) for each velocity map. 
    Below each velocity map is the galaxy identification and emission line fitted. 
    The observed flux maps are shown in column two. 
    White pixels (not shown here) correspond to regions that are outside the MUSE field-of-view or were masked during fitting. 
    The extent of the spatial co-ordinates of the cube used to model the emission and column density of the absorber are given below each residual map. 
    Residual flux maps are found in the third column, where each pixel is coloured by the residual (data $-$ model) normalized by the noise. 
    The absorber and galaxy redshift along with the number of associated galaxies within $\pm 500$ \kms of the absorber are written below the residual map.  
    Associated galaxies that are at the smallest impact parameter to the absorber have a bold border around the plots. 
    }
    \label{fig:Maps1}
\end{figure*}

\begin{table*}
\begin{center}
\caption{\textbf{Summary of kinematic properties for the modelled galaxies.}
The kinematic and stellar properties of galaxies within $\pm 500$ \kms of \ion{H}{i} absorbers (highlighted in bold). 
For each modelled galaxy, the identification and emission line used for modelling is given. 
Velocity differences ($\Delta v$) are calculated with respect to the absorber redshift. 
Star-formation rates (SFRs) and limits are based on \citet{Kennicutt1998}, and we use the H$\alpha$ empirical relation when available. 
At $z \gtrsim 0.4$, we use the [\ion{O}{ii}] $\lambdaup \lambdaup 3726, 3729$ luminosity to estimate the SFR. 
Dust corrections are performed for galaxies with measured H$\beta$ and H$\alpha$ emission-line fluxes.
Stellar mass estimates are derived from SED fitting with \textsc{lephare} using HST broadband imaging \citep[][Augustin et al. in prep.]{LEPHARE}.
The maximum rotational velocity ($V_{\rm max}$), velocity dispersion ($\sigma$) and inclination ($i$) measurements are derived from \textsc{galpak}. 
We adopt the photometric position angle measured using \textsc{galfit} \citep{Peng2002} on the HST photometry (Karki et al. submitted.) when the kinematic PA is not available from \textsc{galpak}. 
The azimuthal angle, $\Phi$, is preferentially calculated using PA$_{\rm kin}$ when available. 
A value of $90$ degrees means that the absorber is orientated along the minor axis of the galaxy. 
Whether the absorber is tracing gas that is part of the disk (Disk), inflowing (In) or outflowing (Out), part of the intragroup medium (IGrM) or ambiguous (A) is given in the last column. 
This label is assigned to only one galaxy in an absorber-galaxy system unless the gas is in the intragroup medium or ambiguous. 
A machine-readable table of associated galaxies and their properties is available online. 
We note that the online table contains both kinematic and photometric inclination and position angle measurements. 
Entries where no measurements are available are filled as $-999$.}
\begin{tabular}{llcccccccccc}
\hline\hline
ID & Line & $b$ & $\Delta v$ & SFR & $\log M_{*}/M_{\odot}$ & $V_{\rm max}$ & $\sigma$ & $i$ & PA & $\Phi$ & Flow \\
    &      &  (kpc) & ($\kms$\!) & (M$_\odot$ yr$^{-1}$) & & (\kms\!) & (\kms\!) & (deg) & (deg) & (deg) &  \\
\hline
\multicolumn{12}{c}{\bolden{Q$0138-0005$, $z_{\rm quasar} = 1.34$, $z_{\rm abs} = 0.7821$, $\logNHIunit = 19.81 \pm 0.08$}} \\
\hline
Q0138m0005\_14 & [\ion{O}{ii}] & 82 & -3.4 & $6.9 \pm 2.4$ & $9.8 \pm 0.2$ & $100 \pm 22$ & $141 \pm 6 $ & $67 \pm 9$ & $100 \pm 6$ & $80 \pm 6$ & A \\
\hline
\multicolumn{12}{c}{\bolden{Q$0152-2001$, $z_{\rm quasar} = 2.06$, $z_{\rm abs} = 0.383$, $\logNHIunit < 18.78$}} \\
\hline
Q0152m2001\_5 & H$\alpha$ & 60 & 84 & $0.73 \pm 0.4$ & $11.25 \pm 0.14$ & $116 \pm 1$ & $14 \pm 2$ & $79 \pm 1$ & $142.1 \pm 0.3$ & $9.3 \pm 0.3$ & In \\
Q0152m2001\_7 & H$\alpha$ & 150 & 350 & $0.18 \pm 0.4$ & $11.00 \pm 0.13$ & $162 \pm 3$ & $4 \pm 4$ & $79 \pm 2$ & $129 \pm 2$ & $58 \pm 2$ & \\
Q0152m2001\_13 & H$\alpha$ & 84 & 330 & $0.10 \pm 0.4$ & $10.5 \pm 0.2$ & $75 \pm 6$ & $5 \pm 4$ & $54 \pm 6$ & $153 \pm 4$ & $6 \pm 4$ & \\
... & ... & ... & ... & ... & ... & ... & ... & ... & ... & ... & ... \\
\hline\hline 				       			 	 
\label{tab:kin_properties}
\end{tabular}			       			 	 
\end{center}			       			 	 
\end{table*}	

\section{Metallicity dependence on galaxy orientation}
Out of the initial 79 galaxies associated with 32 Ly-$\alpha$ absorbers, 48 have kinematic position angle measurements from \textsc{galpak}. 
These measurements are complemented by an additional 19 photometric PA measurements using \textsc{galfit} on the HST broadband imaging (Karki et al. submitted). 
For galaxies with both measurements, the position angles agree within their $1\sigma$ uncertainties. 
To calculate the azimuthal angle ($\Phi$), we preferentially use the kinematic PA when available. 
We then set a restriction on the inclination $i > 30^\circ$ to ensure face-on galaxies with significant errors on the PA are removed from our sample. 
This leaves a total of 56 galaxies with robustly measured azimuthal angles associated with 22 Ly-$\alpha$ absorbers.
The remaining 23 galaxies not considered in this study because they were too faint to be modelled (12) or failed to meet the inclination requirement (11). 
A summary of the sample is shown in \autoref{tab:sample}. 

\begin{table}
\caption{\textbf{Sample summary of modelled galaxies.} 
We consider only the sample of modelled galaxies with inclinations $i > 30^\circ$ to limit potential errors in the position angle measurement. }
\begin{center}
\begin{tabular}{cccc}
    \hline \hline
    Criterion & Tool & Number & \bf{Total} \\
    \hline
    Kinematic modelling & \textsc{galpak} & 48 & \multirow{3}{*}{\bf{79}}\\
    Photometric modelling & \textsc{galfit} &  19 & \\ 
    Not modelled & &  12 & \\
    \hline
    Inclination $> 30^\circ$ & &  56 & \multirow{2}{*}{\bf{67}}\\
    Inclination $< 30^\circ$ & &  11 & \\
    \hline
    \hline
\end{tabular}
\end{center}
\label{tab:sample}
\end{table}

\subsection{Distribution of azimuthal angles}
\begin{figure*}
    \includegraphics[width=\textwidth]{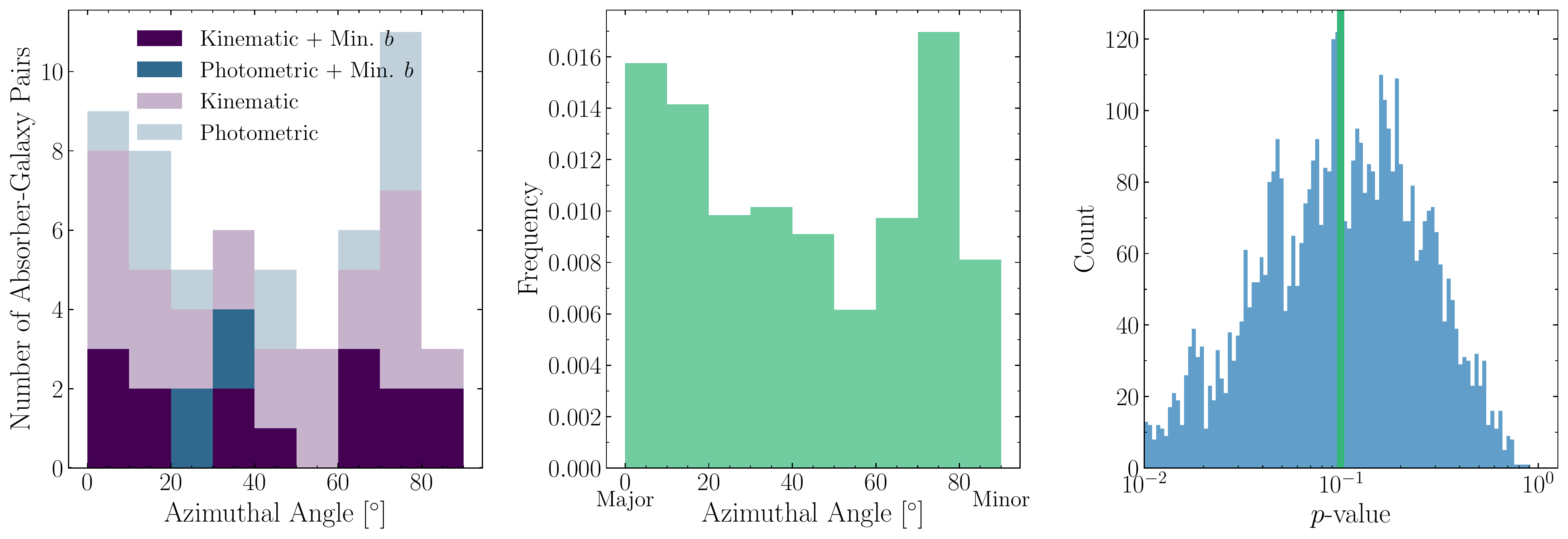}
    \caption{The distribution of azimuthal angles between absorbers and galaxies with inclination $i > 30^\circ$. 
    We set this restriction on the inclination to remove face-on galaxies with large errors in their modelled position angle. 
    In the left histogram, the purple and blue colours respectively represent galaxies with measured kinematic and photometric PAs, derived from modelling the MUSE cubes and HST imaging respectively. 
    Opaque bars represent galaxies found at the closest impact parameter to the QSO sightline, while translucent bars represent PA measurements of the other galaxy counterparts further from the absorber. 
    To test whether the distribution of azimuthal angles is bimodal, we generate 5,000 iterations of the initial distribution by randomly sampling the uncertainties in $\Phi$ and perform the Hartigan's dip test for each iteration \citep{Hartigandiptest}. 
    The final frequency distribution from the 5,000 samples is shown in the middle histogram and the corresponding distribution of $p$-values are shown on the right. 
    The median $p$-value is $\approx 0.1$ (indicated by the vertical green bar) which suggests there is marginal evidence for a bimodal distribution in azimuthal angles. 
    }
    \label{fig:azim_dist}
\end{figure*}

We show the distribution of \ion{H}{i} absorbers around galaxies in \autoref{fig:azim_dist}.
In the left histogram, galaxies with measured kinematic position angles are shown in purple, while the blue represents photometric PAs. 
Galaxies found at closest impact parameter to the absorber are given an opaque bar. 
We impose a restriction on the inclination ($i > 30^\circ$) for the sample to remove face-on galaxies with large errors in their  position angle and by extension, the azimuthal angle. 
This leaves a total of 56 galaxies.  
We test for bimodality by performing the Hartigan's dip test \citep{Hartigandiptest}.  
To account for errors in the azimuthal angle, we repeat the dip test on 5,000 realisations with randomly sampled errors. 
The combined frequency distribution of the 5,000 iterations is depicted in the middle panel of \autoref{fig:azim_dist} and we find that the initial peak in the number count near $\Phi = 75^\circ$ diminishes in significance after accounting for errors in the measurement of $\Phi$. 
The corresponding distribution of $p$-values is displayed in the right panel, where the vertical green bar shows the median $p$-value of the iterations is $\lesssim 0.1$. 
This suggests there is marginal evidence for a bimodality in the azimuthal angle distribution of \ion{H}{i} absorbers around galaxies. 
A larger sample of galaxy-absorber pairs is required to confirm this signal with more certainty. 

\begin{figure*}
    \includegraphics[width=\textwidth]{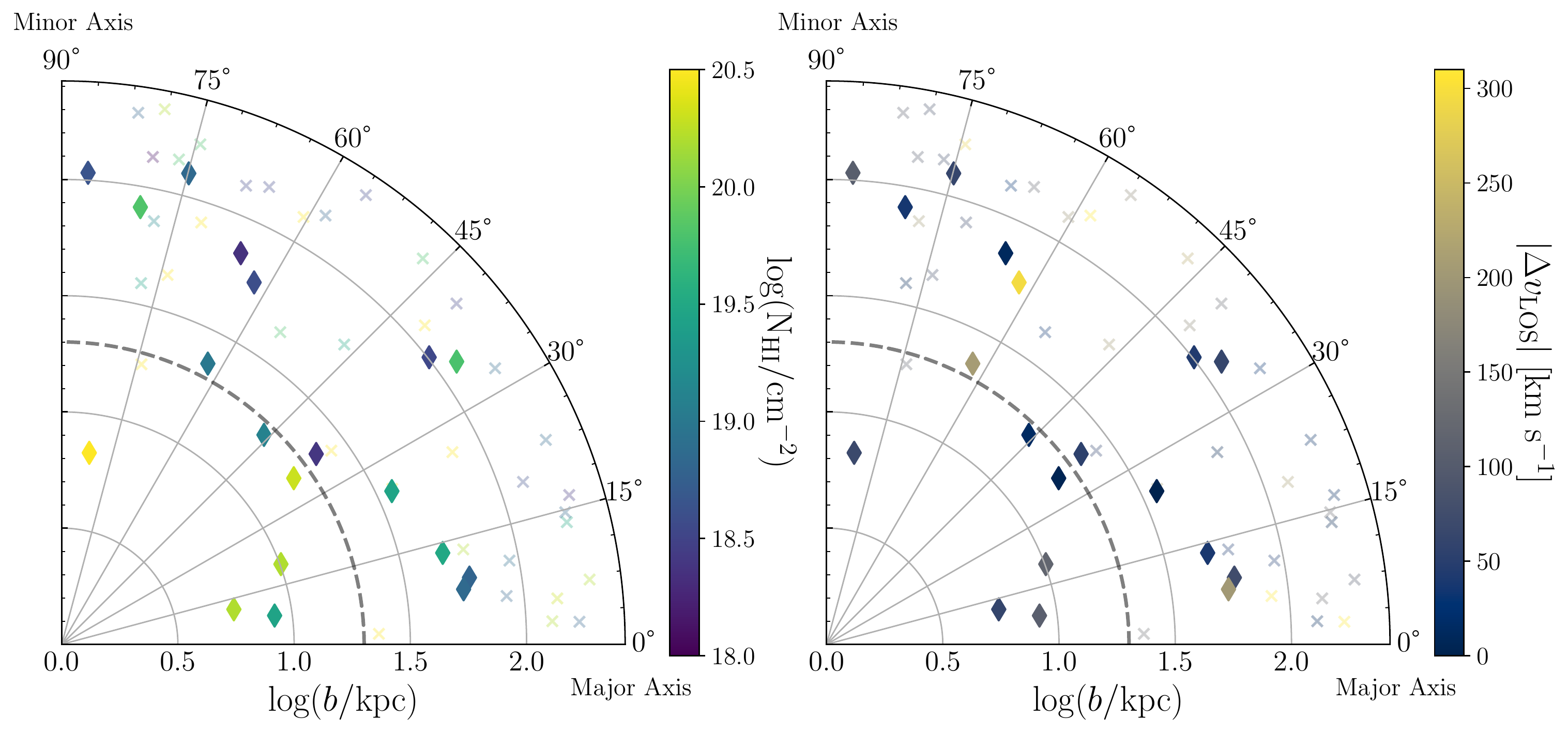}
    \caption{Polar plots illustrating the distribution of the \ion{H}{i} absorbers as a function of azimuthal angle and impact parameter. 
    Diamonds represent galaxies that are found at closest impact parameter to a given absorber and each point is coloured by the \ion{H}{i} column density (left) or the absolute value of the line of sight velocity difference between galaxy and absorber (right). 
    Crosses show the location of associated galaxies at larger impact parameters when multiple galaxies are associated with a single absorber. 
    Points located near $90^\circ$ are found near the galaxy minor axis. 
    The black dashed line marks an impact parameter of $b = 20$ kpc.}
    \label{fig:azim_polar}
\end{figure*}

We show polar plots in \autoref{fig:azim_polar} to visualise the distribution of \ion{H}{i} absorbers more clearly. 
In the left panel, we see the anti-correlation between \ion{H}{i} column density and impact parameter \citep[see][for more analysis of this effect]{Weng2022}. 
There is also little correlation between the line of sight velocity difference between absorber and galaxy, and the polar position, as the median $|\Delta v_{\rm LOS}|$ values in the two bins $0^\circ < \Phi < 30^\circ$ and $60^\circ < \Phi < 90^\circ$ are respectively 75 and 82 \kms\!. 

\subsection{Metallicity dependence on azimuthal angle}

\begin{figure*}
    \includegraphics[width=0.8\textwidth]{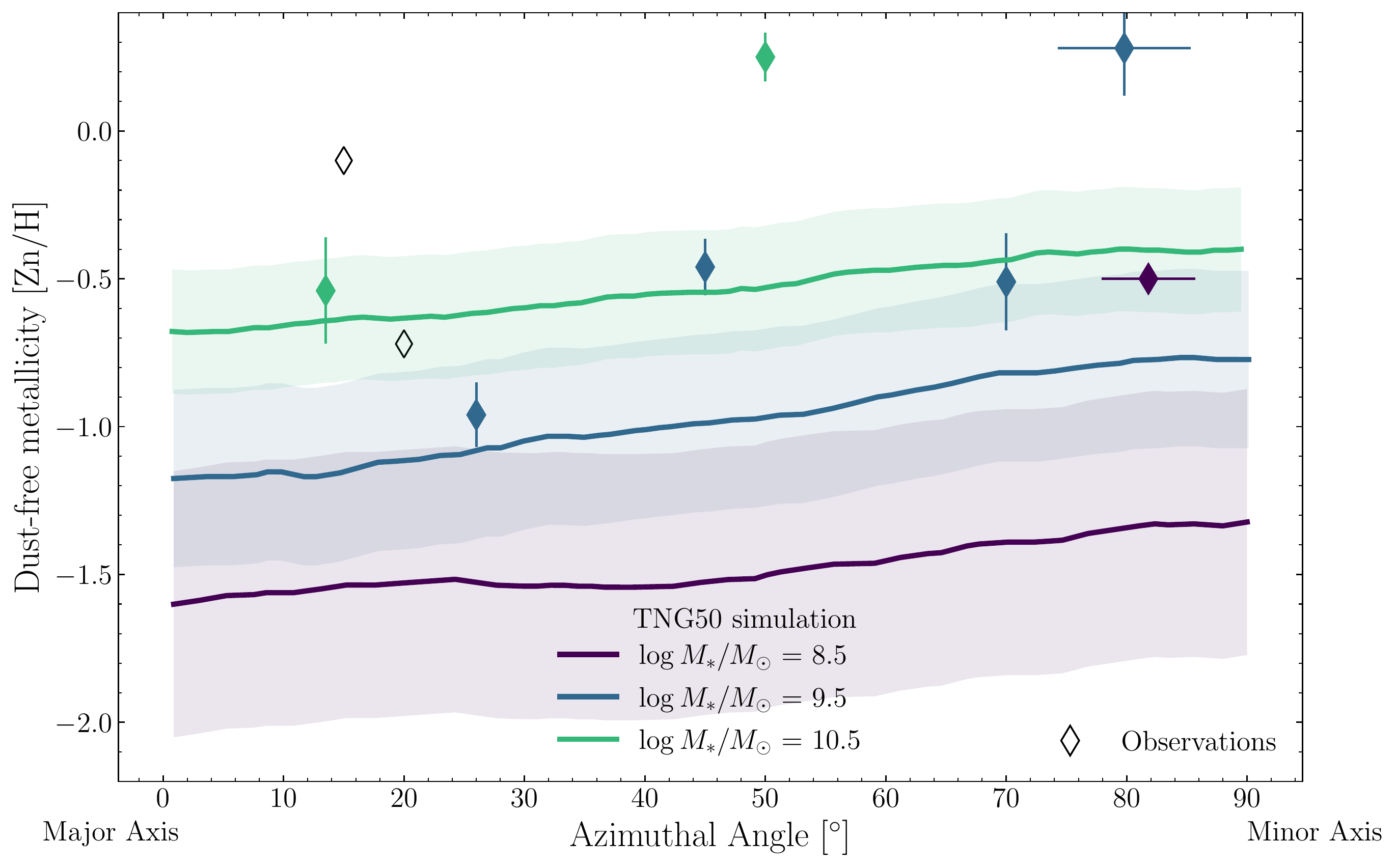}
    \caption{The absorber metallicity as a function of azimuthal angle. 
    Diamonds represent results from this survey and previous works \citep{Peroux2011, Peroux2012, Bouche2013, Bouche2016} that measured metal abundances directly. 
    The coloured lines are the trends predicted by the TNG50 simulation \citep{Peroux2020} for galaxies of different stellar masses at $z = 0.5$ and $b = 100$ kpc. 
    The shaded regions represent a $1\sigma$ deviation. 
    Each galaxy is coloured by their stellar mass corresponding to the coloured lines, while the unfilled black diamonds are galaxies without $M_*$ measurements. 
    At present, there are too few \ion{H}{i} absorbers with robust metallicities and galaxy counterparts to confirm gas near the minor axis is more metal-enriched than gas near $\Phi = 0^\circ$. 
    }
    \label{fig:Z_azim}
\end{figure*}

In \autoref{fig:Z_azim}, we show the absorber metallicity against azimuthal angle using data from previous works \citep{Peroux2011, Peroux2012, Bouche2013, Bouche2016} and the new results presented here from the MUSE-ALMA Halos survey. 
These dust-free metallicities are determined from zinc abundances of dense neutral gas. 
We colour each point by the stellar mass using one dex bins from $\log (M_{*}/M_{\odot}) = 8.0$ to $11.0$. 
Result from the TNG50 simulation \citep{Nelson2019a, Pillepich2019} are plotted for an impact parameter of $b = 100$ kpc, redshift $z = 0.5$ and stellar masses $\log (M_{*}/M_{\odot}) = 8.5$, $9.5$ and $10.5$ \citep{Peroux2020} and the shaded bands represent deviations of $1\sigma$. 
The stellar mass significantly affects the normalisation of the trend, but the metallicity difference between gas near the major and minor axes is approximately 0.3 dex for all stellar mass bins.  
We note that \citet{Peroux2020} find that the relationship diminishes for lower impact parameters ($b < 50$ kpc) and higher redshifts ($z > 1.5$). 
Additionally, limiting the column density to $\logNHIunit > 17.0$ washes out the signal in the simulations because of the limited statistics. 
Hence, it is unsurprising that we find little evidence for a correlation between absorber metallicity and azimuthal angle given we have limited the sample to strong \ion{H}{i} ($\logNHIunit > 19.0$) absorbers with dust-free metallicity measurements. 
Simulations suggest that a sample of $\sim$100 $\logNHIunit > 19.0$ absorbers (where the fraction of ionised gas is expected to be small) with reliable metallicities is required to confirm gas near the minor axis is more metal-enriched than gas near the major axis.

\section{Characterising gas flow origins}
Identifying broader relationships between absorber properties and the galaxy orientation is useful to characterise the global properties of the circumgalactic medium, but identifying gas flows for individual systems enables us to connect galaxy properties with the galactic baryon cycle. 
In this section, we examine the rotation maps of galaxies associated with individual absorbers to infer whether they originate from the galaxy disk, are inflowing or outflowing or arise from other phenomena. 

Previous works have already performed a kinematic analysis of several galaxy-absorber systems in the MUSE-ALMA Halos survey and we summarise it here.
The two absorbers at $z_{\rm abs} = 0.3830$ and $z_{\rm abs} = 0.7802$ towards quasar Q0152$-$0005 were found to respectively trace inflowing and outflowing gas \citep{Rahmani2018a, Rahmani2018b}. 
In a study of the neutral, ionized and molecular gas phases, the $z_{\rm abs} = 0.633$ absorber towards Q0420$-$0127 is found to uncover signatures of outflows or intragroup gas \citep{Klitsch2018}. 
A rich galaxy group is found at $z_{\rm abs} = 0.313$ towards Q1130$-$1449 and the Ly-$\alpha$ absorber appears to trace intragroup gas \citep{Peroux2019}. 
Finally, the absorber towards Q2131$-$1207 is consistent with co-rotating gas accreting onto the galaxy at lowest impact parameter to the absorber \citep{Peroux2017, Szakacs2021}. 
The remaining 22 absorbers with galaxy counterparts are analysed in this work and the final designation for each absorber is noted in \autoref{tab:kin_properties}. 
An individual discussion for each absorber is contained in Appendix \ref{app:flows}.

\subsection{Galaxy disks}
\begin{figure*}
    \includegraphics[width=0.8\textwidth]{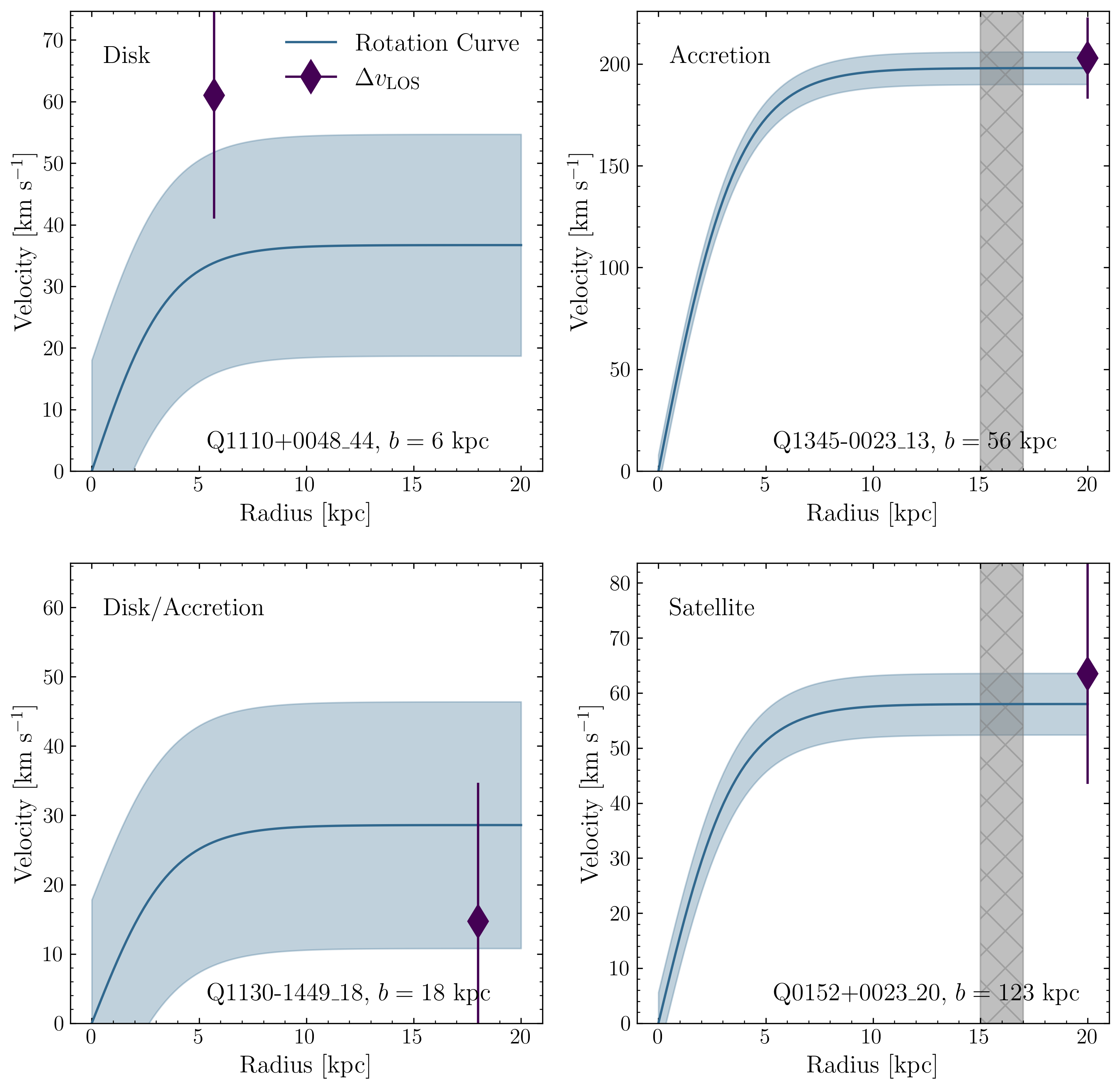}
    \caption{Comparisons between the galaxy rotation curve and absorber line of sight velocities for cases where the disk or gas accretion is being traced. 
    The blue line represents the galaxy rotation curve modelled by a hyperbolic tangent and the shaded region represent the $1\sigma$ error. 
    We plot the velocity of the absorber relative to the systemic redshift of the galaxy as a purple diamond. 
    A vertical hatched region is used to illustrate that the impact parameter of the absorber is beyond the limits of the $x$-axis. 
    We provide the gas flow origin in text at the top-left, while the galaxy ID and absorber impact parameter are found at the bottom of each plot. 
    In three cases, we find the absorber velocity is consistent with the rotating disk or co-rotating halo of the galaxy at nearest impact parameter. 
    For the bottom-right case, the absorber is beyond the virial radius of the nearest galaxy and likely arises from a faint, quiescent galaxy. 
    }
    \label{fig:rotcurve}
\end{figure*}

We find one case where the absorber evidently intersects the galaxy disk. 
Such absorbers are characterised by a strong \ion{H}{i} absorber ($\logNHIunit > 20.0$) found within $\sim$20 kpc of a galaxy \citep{Wolfe1986, Zwaan2005, Peroux2005}. 
The clearest example of disk gas being probed is the absorber towards Q1110$+$0048 at $z_{\rm abs} = 0.5604$ where the host galaxy is found at an impact parameter of $6$ kpc. 
We find the $\logNHIunit = 20.2$ absorber at the expected velocity sign and magnitude as the modelled rotational map (see top left panel of \autoref{fig:rotcurve}). 
For the remaining three cases, no kinematic modelling is possible for the galaxies found at impact parameters $b < 20$ kpc. 
Nevertheless, we also attribute these absorbers to arise from the galaxy disk given their column densities ($\logNHIunit > 20.0$) and low impact parameters. 
The $|\Delta v_{\rm LOS}|$ values for these absorbers range from 5 to 150 \kms and are consistent with typical rotation speeds.  
We calculate the azimuthal angle between the QSO sightline and galaxy using the HST photometry to range from $20^\circ$ to $40 ^\circ$. 
Thus, we do not find the absorbers near the minor axes where outflowing gas may be expected. 
While we lack direct evidence the absorber is rotating with the galaxy disk without velocity maps, damped Lyman-$\alpha$ absorbers have long been associated with the rotating disk of galaxies \citep{Wolfe1986, Wolfe2005}. 
More recent simulations predict a covering fraction of 50 per cent for DLAs within 0.1 $R_{\rm vir}$ (15 kpc) of galaxies with halo mass $10^{12}$ M$_\odot$ at $z \sim 0$ \citep{Stern2021}. 
These findings support the idea that the strong \ion{H}{i} absorbers at $z < 1$ in this study arise from galaxy disks, whereas DLAs at $z \gtrsim 2$ increasingly probe the inner circumgalactic medium. 

\subsection{Accretion}
Similar to absorbers that probe the galaxy disk, gas that is co-rotating with a galaxy can be detected by comparing the velocity of the absorber with the ionized gas velocity field. 
The difference is that gas accreting onto the galaxy is lower column density and can be found at larger impact parameters. 
We find that the $z_{\rm abs} = 0.6057$ absorber towards Q1345$-$0035, where the host galaxy is found at an impact parameter of $56$ kpc, is consistent with infalling gas. 
The $\logNHIunit = 18.85$ absorber is found with the same velocity sign and magnitude to the modelled rotational map (bottom right panel of \autoref{fig:rotcurve}) and is likely tracing gas in the circumgalactic medium co-rotating with the galaxy disk. 

Less certain cases include the $\logNHIunit < 19.1$ absorber towards Q1130$-$1449 at $z_{\rm abs} = 0.1906$. 
The impact parameter of 18 kpc suggests the gas originates from the galaxy disk, but the upper limit on the \ion{H}{i} column density signifies the gas is perhaps accreting. 
We also find dense gas consistent with co-rotation for the absorber towards Q0152$+$0023 at $z_{\rm abs} = 0.4818$ (top right panel of \autoref{fig:rotcurve}). 
However, it is unclear whether such cool dense gas ($\logNHIunit = 19.78$) can be found co-rotating at impact parameters $>100$ kpc. 
Importantly, the stellar mass of the galaxy Q0152$+$0023\_20 associated with the absorber is $\log(M_*/M_\odot) = 8.1 \pm 0.1$ and the estimated virial radius, $R_{\rm vir}$, is $\sim$70 kpc \citep{Puebla2017}. 
Despite the alignment in velocity, the absorber is beyond the halo of the associated galaxy and it is possible that a passive, low-mass galaxy (SFR $< 0.19\ \Moyr\!$, $\log(M_*/M_\odot) < 8.5$) below the MUSE detection threshold nearer the QSO sightline hosts the dense absorber. 


Within the full MUSE-ALMA Halos Survey sample of \ion{H}{i} absorbers, we find only three cases that are consistent with gas accretion. 
Cold streams in the circumgalactic medium are expected to be observed in absorption because of their high densities and temperatures $\sim$10$^4$ K \citep{Fumagalli2011b, Faucher2015, Hafen2017}. 
Simulations predict that dense \ion{H}{i} absorbers at $z \gtrsim 3$ trace cold-mode accretion \citep{vandevoort2012} and detections of low-metallicity Lyman-limit systems (LLSs) have been suggested to trace these inflows \citep{Ribaudo2011, Fumagalli2011a, Lehner2013}. 
However, at lower redshifts, direct detections of gas accretion are sparse for both down-the-barrel and transverse absorption-line studies \citep{Kacprzak2010, Martin2012, Rubin2012, Ho2017, Zabl2019}. 
Our results echo the findings of studies without pre-selection of targets and find only $\sim$10 per cent of \ion{H}{i} absorbers consistent with accretion. 

\subsection{Outflows}
In total, we find seven absorbers that are within $30^\circ$ of the projected minor axis of nearby galaxies. 
However, only three out of these seven absorbers can be confidently attributed to gas that is outflowing. 
One clear case of outflowing gas is seen the $z_{\rm abs} = 0.7869$ absorber towards in Q1554$-$203. 
The absorber is found at an impact parameter of $23$ kpc and azimuthal angle of $62^\circ$ (see row 2 in \autoref{fig:Maps10}). 
The gas is unlikely to be inflowing or associated with the galaxy disk as the velocity of the absorber is opposite in sign to the ionized gas velocity field. 
A line of sight velocity difference between the galaxy and absorber of $\sim$200 \kms is also typical of outflows. 
The absorber towards Q0454$+$039 at $z_{\rm abs} = 1.1532$ with column density $\logNHIunit = 18.59$ aligns with the minor axis ($\Phi = 62^\circ$) of the nearest galaxy (see second row of \autoref{fig:Maps4}). 
At an impact parameter of 60 kpc and line of sight velocity difference of $-290$ \kms, the gas is likely tracing neutral gas entrained in an outflow. 
While the absorber and ionized gas are both blueshifted with respect to the galaxy systemic redshift, there is a $> 250$ \kms discrepancy between the maximum rotational velocity of the galaxy and the absorber velocity. 
Finally, we find the DLA at $z_{\rm abs} = 0.3950$ towards Q1229$-$021 is consistent with cold gas entrained in an outflow. 
At an impact parameter of $6$ kpc and azimuthal angle of $81^\circ$, the neutral gas velocity is inconsistent with the ionized gas disk. 
We measure a line of sight velocity difference of $70$ \kms\!, which suggests the gas is still accelerating at this small distance from the galaxy. 

The remaining four absorbers that align with the minor axis possess unclear gas flow origins. 
The uncertainty arises from two issues: the absorber velocity is inconsistent with outflowing gas, or there are multiple galaxies at similar impact parameters to the absorber. 
Addressing the former concern first, we find that the absorber towards Q0138$-$0005 at $z_{\rm abs} = 0.7821$ has only one galaxy counterpart at an impact parameter of 80 kpc (first row of \autoref{fig:Maps1}). 
The absorber is aligned with the minor axis of the associated galaxy and has a zinc abundance of $\rm{[Zn/H]} = 0.28 \pm 0.16$. 
While both these properties are consistent with metal-enriched gas expelled by the single galaxy counterpart, the absorber velocity is $< 40$ \kms from the galaxy systemic redshift. 
Another similar example is the absorber towards Q1229$-$021 at redshift $z_{\rm abs} = 0.7572$ (last row in \autoref{fig:Maps7}) which is separated by only $15$ \kms from the redshift of its galaxy counterpart. 
A possible explanation for these low velocities is that there is a large velocity component orthogonal to the sightline as these galaxies have inclinations $i > 65^\circ$. 
The line of sight velocity is not necessarily well-correlated with the radial velocity of the gas. 
Alternatively, the neutral gas may be static and not co-rotating with the halo. 

For the two absorbers at $z_{\rm abs} = 0.7691$ and $0.8311$ towards Q1229$-$021, there are at least three galaxies found at the absorber redshift and at similar distances from the QSO sightline. 
It becomes difficult to determine the gas flow origin of the absorber as gravitational interactions between the galaxies can cause gas to be stripped. 
The individual cases are discussed in Appendix \ref{app:flows} and a more detailed exploration of intragroup gas is explored in the following subsection. 
We choose to label these four unclear cases as possible outflows.

\subsubsection{Does the gas escape?}
We can determine whether the neutral gas escapes the galaxy halo by comparing the absorber velocity with the escape velocity ($V_{\rm esc}$). 
The escape velocity at a given radius, $r$, is calculated assuming a singular isothermal sphere \citep{Veilleux2005}:
\begin{equation}
\centering
    V_{\rm esc} = V_{\rm vir} \times \sqrt{2 \left(1 + \ln{\frac{R_{\rm vir}}{r}}\right)},
\end{equation}
where $V_{\rm vir}$ and $R_{\rm vir}$ are the virial velocity and radius respectively. 
Here, we assume the radius to be the impact parameter ($r \approx b$). 
We estimate $V_{\rm vir}$ using the prescription in \citet{Schroetter2019} where $V_{\rm vir} \approx 1.2 \times S_{0.5}$. 
Here, $S_{0.5} = \sqrt{0.5 \times V_{\rm max}^2 + \sigma^2}$ is the kinematic estimator and is a function of the rotational velocity and velocity dispersion, $\sigma$ \citep{Weiner2006}. 
Using $V_{\rm vir}$, we can then approximate the virial radius to be $R_{\rm vir} \approx V_{\rm vir}/10H(z)$ where $H(z)$ is the Hubble constant at redshift $z$. 
The estimated escape velocity values are tabulated in \autoref{tab:escape}. 

\begin{table}
\caption{\textbf{Comparison of line of sight absorber velocity with the escape velocity for galaxy-absorber pairs consistent with outflows.} 
The line of sight velocity of the absorber relative to the galaxy is roughly equal to the escape velocity assuming a singular isothermal sphere. 
Given that $\Delta v_{\rm LOS}$ does not account for velocities orthogonal to the line of sight, it is possible that the \ion{H}{i} absorbers trace neutral gas escaping the potential of the galaxy. 
While $|\Delta v_{\rm LOS}|$ < $V_{\rm esc}$, the radial velocity of the gas may be larger than the escape velocity. }
\begin{center}
\begin{tabular}{cccccc}
    \hline \hline
    QSO & $z_{\rm abs}$ & \logNHI & $b$ & $|\Delta v_{\rm LOS}|$ & $V_{\rm esc}$ \\
     &  & $\log(\rm cm^{-2})$ & kpc & \kms\! & \kms\! \\
    \hline
    Q0454$+$039 & 1.1532 & 18.59 & 60 & 290 &  300 \\
    Q1229$-$021 & 0.3950 & 20.75 & 6 & 70 & 80 \\ 
    Q1554$-$203 & 0.7869 & <19.0 & 23 & 210 & 270  \\
    \hline
    \hline
\end{tabular}
\end{center}
\label{tab:escape}
\end{table}

While the absorber velocity relative to the galaxy is less than the escape velocity for all three likely cases of outflows, we do not take into account the velocity component orthogonal to the line of sight. 
Hence, it is possible that the radial velocity of the outflowing neutral gas exceeds the escape velocity and will be ejected from the galaxy halo. 

\subsection{Alternative phenomena}
Beyond the absorber origins discussed earlier in this section, there are other phenomena that may produce \ion{H}{i} absorption around galaxies. 
In particular, recent studies using MUSE reveal that roughly 50 per cent of Ly-$\alpha$ absorbers have multiple galaxies within a velocity window of $\pm 500$ \kms \citep{Chen2020, Hamanowicz2020, Weng2022, Berg2023}. 
This suggests the absorption may arise from the intragroup medium between galaxies \citep{Gauthier2013, Nielsen2018, Dutta2023}. 
It is difficult to identify intragroup gas because we require kinematic modelling for all galaxies in the overdensity to exclude gas flows. 
The two galaxies associated with the $z_{\rm abs} = 0.3283$ absorber towards Q1130$-$1449 are roughly equidistant in projected distance from the absorber (75 and 90 kpc). 
There is no hint of inflows or outflows from the relative velocity and geometry of the absorber. 
Instead, we find that the velocity of the absorber is between the two galaxy systemic redshifts which suggests we may be tracing intragroup gas with \logNHIunit < 18.9. 
There are no other clear examples of absorbers tracing gas between galaxies, but there are several cases where multiple explanations are viable such as the two absorbers towards Q1229$-$021 discussed in the previous section.

Another important consideration is that low-mass, quiescent galaxies or satellites hosting the absorber may not be detected in our observations. 
Indeed, there is a possible example of such a case with the four galaxies at impact parameters of 122 to 190 kpc associated with the absorber towards Q0152$+$0023. 
We previously discussed in this section whether the gas may arise from accretion due to an alignment between the rotational velocity and absorber velocity (top right panel of \autoref{fig:rotcurve}), but we noted the absorber is beyond the virial radius of the nearest galaxy. 
In fact, the $\logNHIunit = 19.78$ absorber is beyond the virial radii of all four galaxies
This points to the hypothesis that there is a galaxy with stellar mass $\log(M_*/M_\odot) < 8.6$ and SFR $<0.24$ \Moyr below the detection threshold in the MUSE data near the strong \ion{H}{i} absorber \citep{Weng2022}. 
While recent works suggest that the CGM extends beyond the virial radius \citep{Wilde2021, Wilde2023}, it is unlikely to find such dense absorbers with column densities $\logNHIunit \approx 20$ at such large distances from galaxies. 

A final point is that the line of sight velocity is not necessarily a good predictor of the physical line of sight distance between an absorber-galaxy pair. 
In fact, simulations have shown that absorbers can be found at $> 1$ pMpc when applying a velocity cut of $|\Delta v_{\rm LOS}| < 500$ \kms \citep{Rahmati2015, Ho2020, Ho2021}. 
The velocity difference between absorber and galaxy is influenced by peculiar motions of the gas and the Hubble flow at larger separations.  
This suggests that there is always the possibility of chance associations between galaxies and absorbers. 
Quantifying this probability is the goal of an upcoming work using the TNG50 simulations.

\section{Discussion}
In this section, we discuss the incidence rate of the various gas flows in the MUSE-ALMA Halos survey. 
We also examine the relationships between measured galaxy properties and the origin of the gas. 
Additionally, we discuss the limitations of using simple geometric arguments to distinguish outflows and inflows and future improvements to this analysis by studying individual gas components. 

\subsection{The origins of gas in the CGM}
The \ion{H}{i} Ly-$\alpha$ absorbers in the MUSE-ALMA Halos survey appear to trace various phenomena. 
In summary, out of the 32 absorbers in the survey, 27 are found to have at least one galaxy within $\pm 500$ \kms\!. 
The 27 absorbers are comprised of four absorbers tracing the galaxy disk, three tracing accretion, four tracing outflows, two tracing gas in the intragroup medium and one likely tracing an undetected, low-mass galaxy. 
Six absorbers may arise from multiple phenomena while the remaining five (two) have associated galaxies without kinematic modelling (with inclinations $i < 30^\circ$). 
While this sample of absorbers only provides limited statistics, we do find that the accretion of gas onto galaxies is difficult to trace. 
Whether this is caused by the criteria used to identify accretion or an intrinsic property of the accretion itself \citep{Faucher2011} is unclear, but the percentage of accreting absorbers in this work is similar to down-the-barrel studies \citep{Martin2012, Rubin2012} and transverse absorption-line works using \ion{Mg}{ii} \citep{Zabl2019}. 
We find four convincing cases of \ion{H}{i} outflows which is proportionally far less than other surveys \citep[e.g.][]{Schroetter2019}. 
However, we note that there are six other cases where the absorber kinematics and geometry are consistent with outflows, where the interpretation is unclear because of possible interactions with other galaxies in the field or the modelled inclination not meeting our threshold ($i < 30^\circ$). 
Ultimately, we have found that strong \ion{H}{i} Ly-$\alpha$ absorbers probe gas in a variety of environments and of many origins in the complex circumgalactic medium. 
In a future work using the TNG50 simulation, we will address whether the various gas origins discussed here can be distinguished from each other using the available observables of impact parameter, metallicity, line of sight velocity difference and azimuthal angle. 

\subsection{Gas flows and galaxy properties}
With our physical interpreations of the absorber, we examine the properties of the galaxies associated with these absorbers. 
In \autoref{fig:SFR_Mass_flows}, we show the stellar masses and star-formation rates of galaxies associated with the absorbers of various origin. 
Diamonds represent absorbers that likely intersect the galaxy disk, stars and squares represent outflows and inflows respectively, while crosses signify there are multiple origins for the gas. 
For both the stellar mass and SFR, a $\sim$3 dex range of values is observed. 

Inflows are found to be associated with galaxies with large stellar masses $10^{10}$ to $10^{11}$ $M_\odot$, but more photometry is required to estimate $M_*$ for the other galaxies. 
While the star-formation rate of galaxies associated with the various gas flows span two dex, a more important indicator is the star-formation rate per unit area ($\Sigma_{\rm SFR}$). 
Two of the three galaxies where the absorber likely traces outflows have a $\Sigma_{\rm SFR} > 0.1$ M$_\odot$ yr$^{-1}$ kpc${-2}$ \citep{Heckman2002, Heckman2003}. 
The third galaxy does not meet the threshold in $\Sigma_{\rm SFR}$, but we note that the SFR is not corrected for dust and the starburst responsible for the galactic wind might be more localised. 
Currently, a larger sample is required to connect the gas flows traced by absorbers to the galaxies that host the neutral gas. 


\begin{figure*}
    \includegraphics[width=0.95\textwidth]{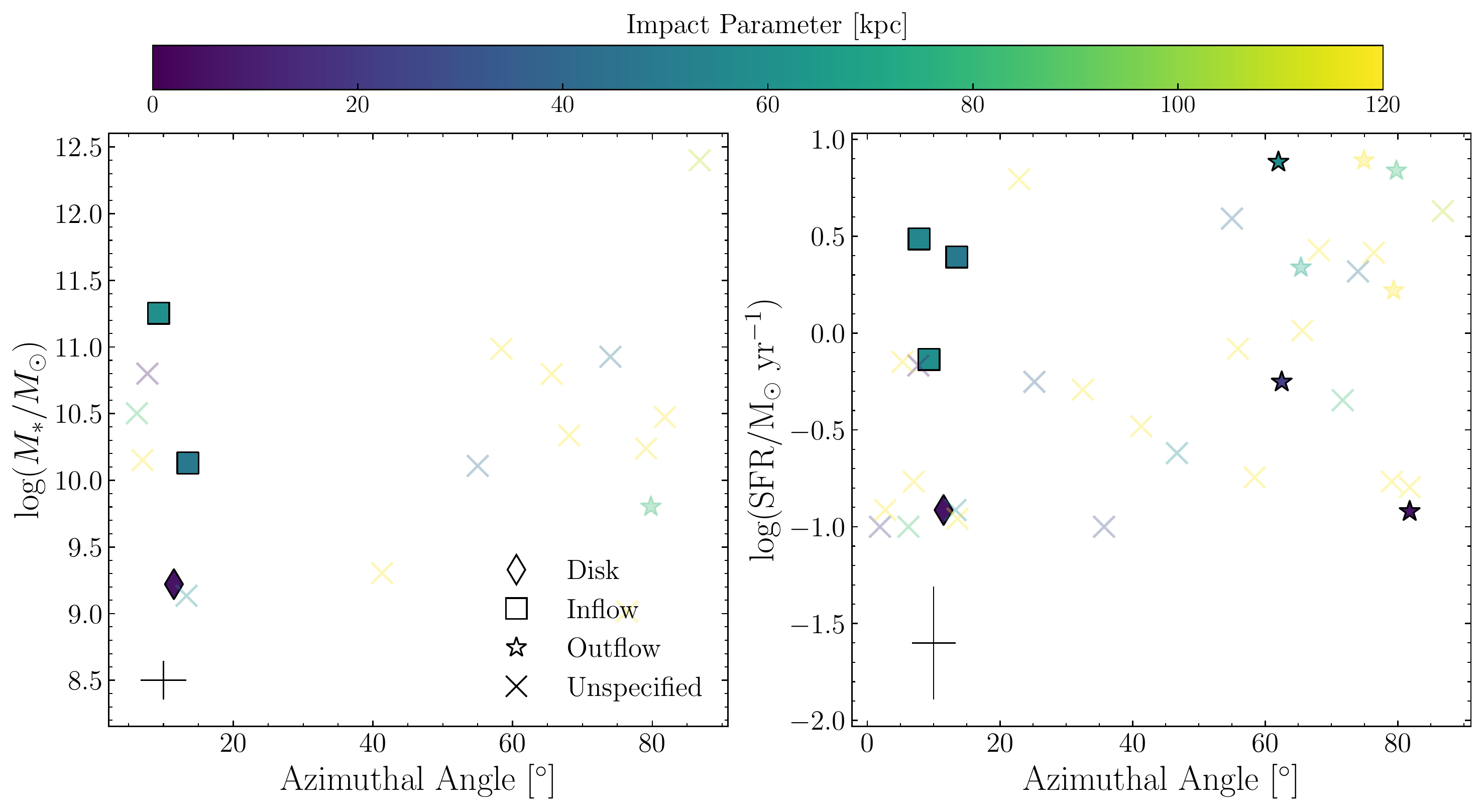}
    \caption{The stellar properties of the galaxies associated with an absorber. 
    The left plot shows the stellar mass of galaxies associated with absorbers at a given azimuthal angle. 
    Stellar masses are derived from spectral energy distribution (SED) fitting of the HST broadband imaging (Augustin et al. in prep). 
    Different symbols represent galaxy-absorber pairs that are found to trace inflowing, outflowing or gas in the disk while faint crosses represent pairs with ambiguous origins. 
    Symbols with a black border are cases where the gas origin has been confidently identified, while those with lower transparency are possible cases. 
    Each galaxy is coloured by the impact parameter from the absorber. 
    We find absorbers are associated with galaxies that span four dex in stellar mass.
    On the right, we show the dust-uncorrected star-formation rate of galaxies measured using the H$\alpha$ or [\ion{O}{ii}] emission lines \citep{Weng2022}. 
    We find that the galaxies associated with inflows and outflows do not differ significantly in their SFRs. 
    The median errors in the plotted properties are shown as a cross in the bottom left of both plots. 
    }
    \label{fig:SFR_Mass_flows}
\end{figure*}

A recent study of galaxies associated with \ion{Mg}{ii} absorbers finds that galaxies with large inflow rates are located above the SFR-$M_*$ main sequence \citep{Langan2023}. 
Using the three galaxies associated with inflows that have both SFR and $M_*$ measurements, we find no such signal and most of the associated galaxies lie on the SFR-$M_*$ main sequence (Karki et al. submitted). 

\subsection{The azimuthal distribution of \ion{H}{i} and metals in the CGM}
Outflows of gas driven by supernovae or AGN are expected to be preferentially aligned with the minor axis and collimate into a biconical shape \citep{Veilleux2005}. 
Both hot- and cold-mode accretion is expected to channel cool gas onto galaxies via cool flows that align with the disk \citep[e.g.][]{Hafen2022, Keres2005}. 
In absorption-line studies of ions such as \ion{Mg}{ii} and \ion{O}{vi} in the CGM, the inflow-outflow dichotomy manifests in the bimodal distribution of azimuthal angles between the quasar line of sight and nearby associated galaxy \citep[e.g.][]{Bouche2012, Kacprzak2015}. 
In this work, we also find marginal evidence for a bimodality in the distribution of \ion{H}{i} absorbers. 
In principle, this bimodality in $\Phi$ should extend to metallicity measurements as outflowing gas is typically more metal-enriched than gas accreting onto a galaxy. 
However, the dependence of metallicity on azimuthal angle is far less clear from observations \citep{Peroux2016, Kacprzak2019, Pointon2019, Wendt2021}. 
Here, we find very little evidence for a correlation between absorber metallicity and azimuthal angle for three different stellar mass bins, and a larger and more homogeneous sample is required to determine whether a relationship exists.

The reasons for the absence of a distinct signal are manifold. 
The trend of gas-phase metallicity versus azimuthal angle depends on properties such as the impact parameter, \ion{H}{i} column density and most significantly, the stellar mass \citep{Peroux2020}.  
These parameters not only affect the normalisation of the signal, but also the magnitude of the discrepancy between low and high azimuthal angles. 
A larger sample of absorbers reaching column densities $\logNHIunit = 13.0$ that are found at impact parameters larger than 50 kpc from their host galaxies that have measured stellar masses is required to fully test the azimuthal dependence of metallicity in the CGM. 
Such a sample will be difficult to construct because of the challenge in measuring gas-phase metallicity due to the uncertainties surrounding photoionization and dust modelling. 
Furthermore, there is the intrinsic inhomogeneity of gas properties at small spatial scales in the CGM that will require large samples to take into account. 
Estimates of line of sight cloud sizes range from sub-parsec to $\sim$100 parsec from ionization modelling \citep[e.g.][]{Churchill2003, Werk2014} and studies find significant ranges in metallicity for different components along a single line of sight \citep[e.g.][]{Zahedy2019, Nielsen2022}, suggesting metals may be poorly mixed in the CGM \citep{Peroux2018, Tejos2021}. 
This is captured by the TNG50 predictions depicted in \autoref{fig:Z_azim}, where the $1\sigma$ errors range $\sim$1 dex in metallicity. 
Moreover, recent work from \citet{Berg2023} suggest there is a population of low-metallicity absorbers residing in overdense regions away from galaxy halos. 
Indeed, the baryon cycle is more complex than a linear combination of gas being expelled out via the minor axis and accreting along the major axis; these processes interact to form the complex, multi-phase circumgalactic medium.

\subsection{The fidelity of geometric and kinematic arguments}
In this work, we adopt an opening angle of $60^\circ$ (corresponding to $\pm 30^\circ$ from the minor axis) to identify gas that is being expelled \citep{Chen2010, Lan2014}, but a diverse range of opening angles have been observed in the literature \citep[e.g.][]{Veilleux2001} and these angles differ depending on the observed wavelength \citep[e.g. see the opening angles for the Circinus galaxy ranging from $15^\circ$ to $100^\circ$][]{Harnett1990, Elmouttie1995, Veilleux1997, Curran1999}. 
It is unclear whether several of the absorbers with azimuthal angles just below the threshold should be considered outflowing. 
Similarly, while inflowing gas from hot- and cold-mode accretion is expected to align in angular momentum with the disk, there has been growing evidence for the condensation of ambient gas in the halo caused by interactions between gas ejected by stellar winds and hot coronal gas \citep{Marinacci2010, Fraternali2017}.
Known as the galactic fountain, the phenomenon is expected to cause neutral gas to `rain' down onto the galaxy disk rather than align with the major axis. 
There is mounting evidence for galactic fountains in the Milky Way, with the kinematics of high- and intermediate velocity clouds consistent with a mixture of outflowing and diffuse inflowing gas \citep{Lehner2022b, Marasco2022}. 
Beyond the local Universe, signatures of fountains flows have been found to persists at impact parameters $b > 5$ kpc \citep{Rubin2022}.
Hence, the absorbers in this sample may also trace this process at $z \sim 0.5$. 

Another consideration is that the strong \ion{H}{i} Ly-$\alpha$ absorbers in this sample are comprised of multiple components at varying velocities to the DLA redshift. 
Recently, photoionization modelling of individual components within \ion{H}{i} systems has been performed to determine cloud properties such as temperature and size \citep{Cooper2021, Zahedy2021, Nielsen2022}. 
The individual clouds and their properties have then been related to outflows, inflows or intragroup gas. 
We leave the modelling of the various metal-line components and estimations of the typical cloud properties embedded in outflowing and inflowing gas to future works.   

\section{Summary and Conclusion}
The MUSE-ALMA Halos survey combines multi-wavelength observations of galaxies associated with 32 $\logNHIunit > 18.0$ absorbers. 
In this work, we have modelled the ionized gas kinematics of 48 galaxies associated these absorbers using the forward modelling algorithm \textsc{galpak} to extract properties such as the rotational velocity, velocity dispersion, inclination and position angle. 
By determining the position and geometry of the absorber with respect to the modelled galaxies, we seek to determine the distribution of gas in the circumgalactic medium and identify the possible origins of these strong \ion{H}{i} absorbers. 
To summarise, we find:

\begin{enumerate}
    \item 
    An excess of absorption sightlines passing near the major and minor axes of galaxies. 
    There is marginal evidence for a bimodal distribution in azimuthal angles between galaxy and absorber after performing the Hartigan's dip test on 5,000 iterations of the data by randomly sampling the errors ($p \approx 0.1$). 
    This is similar to previous studies of the \ion{Mg}{ii} and \ion{O}{vi} ions, suggesting inflows and outflows of gas in the CGM can also be traced by neutral hydrogen. 
    \item 
    That there is little evidence for a dependency of the metallicity on the azimuthal angle for the absorbers in the MUSE-ALMA Halos survey as predicted by simulations. 
    This suggests that gas in the circumgalactic medium is not merely a linear combination of metal-poor inflows and metal-enriched outflows and that other phenomena such as gas recycling and poor metal-mixing are significant. 
    The results from simulations also show that the scatter in metallicities at any given azimuthal angle is comparable to the actual metallicity discrepancy at the minor and major axes. 
    At this stage, simulation results \citep{Peroux2020} advocate that a larger sample of $\sim$100 strong \ion{H}{i} absorbers with dust-free metallicity measurements is still required to recover any signal predicted by simulations.
    \item 
    That \ion{H}{i} absorbers have a variety of origins in the CGM. 
    Absorbers with column densities $\logNHIunit > 20.0$ at impact parameters of $b < 20$ kpc from the nearest galaxy are considered associated with the galaxy disk. 
    Only 15 per cent of absorbers are found to trace the disk, suggesting other processes must account for the remaining absorbers. 
    We find that roughly 10 per cent of absorbers are co-rotating with the halo out to distances up to $60$ kpc and these are suspected to trace gas accretion. 
    The rarity of such cases is in line with previous works. 
    Up to $\approx$30 per cent of absorbers are found within $\pm 30^\circ$ of the minor axis of galaxies and are consistent with outflows, but we only identify four clear cases in the sample. 
    The remaining absorbers trace gas in the intragroup medium, low-mass galaxies below the detection limit of the MUSE data or do not have sufficient data to uncover a physical origin. 
\end{enumerate}

In the future, larger surveys such as the ByCycle survey (PI: P\'{e}roux) using the 4MOST instrument will enable us to compare the kinematics of galaxies with absorbers on a much larger scale \citep{4MOST, 4MOSTcomm}. 
The combination of high-resolution ($R \sim 20,000$) background QSO spectroscopy with deep and complete foreground galaxy surveys \citep{Driver2019} will enable better constraints on how the absorber metallicity varies with azimuthal angle (Szakacs et al., submitted). 
While we have identified the various gas flows traced by dense \ion{H}{i} absorbers, the future modelling of individual gas components will provide information on the gas properties of inflows and outflows (e.g. temperature, density and cloud size). 

The proliferation of absorber follow-up surveys that use integral field spectroscopy has led to a re-characterisation of the physical processes that absorbers trace in the circumgalactic medium and how gas is distributed with respect to galaxies.  
In this work, we find and emphasise that even strong Ly-$\alpha$ absorbers trace gas flows in the circumgalactic medium. 
The fact that dense, neutral gas required for star-formation is seen accreting and being expelled highlights these processes have significant impacts on galaxy evolution and that the CGM plays an important role in regulating the baryon cycle. 

\section*{Acknowledgements}
This research is supported by an Australian Government Research Training Program (RTP) Scholarship.
EMS, GK and SW acknowledge the financial support of the Australian Research Council through grant CE170100013 (ASTRO3D).
VPK and AK acknowledge partial support for GO program 15939 (PI: Peroux) provided through a grant from the STScI under NASA contract NAS5-26555 and NASA grant 80NSSC20K0887 (PI: Kulkarni). 
VPK also gratefully acknowledges additional support from the National Science Foundation grants AST/2007538 and  AST/2009811 (PI: Kulkarni). 
DN acknowledges funding from the Deutsche Forschungsgemeinschaft (DFG) through an Emmy Noether Research Group (grant number NE 2441/1-1).
This research was supported by the International Space Science Institute (ISSI) in Bern, through ISSI International Team project \#564 (The Cosmic Baryon Cycle from Space).

This research also made use of several \textsc{python} packages: \textsc{astropy} \citep{astropy:2013, astropy:2018}, \textsc{matplotlib} \citep{Hunter:2007} and \textsc{numpy} \citep{harris2020array}.

\section*{Data Availability}
Data directly related to this publication and its figures are available upon request. 
The catalogues for the MUSE-ALMA Halos survey have been made public with the publication of \citet{Peroux2022}. 
The raw data can be downloaded from the public archives with the respective project codes.



\bibliographystyle{mnras}
\bibliography{MAH_XI} 

\newpage



\appendix

\section{Additional Velocity Maps}
\label{app:maps}

\begin{figure*}
    \includegraphics[width=0.81\textwidth]{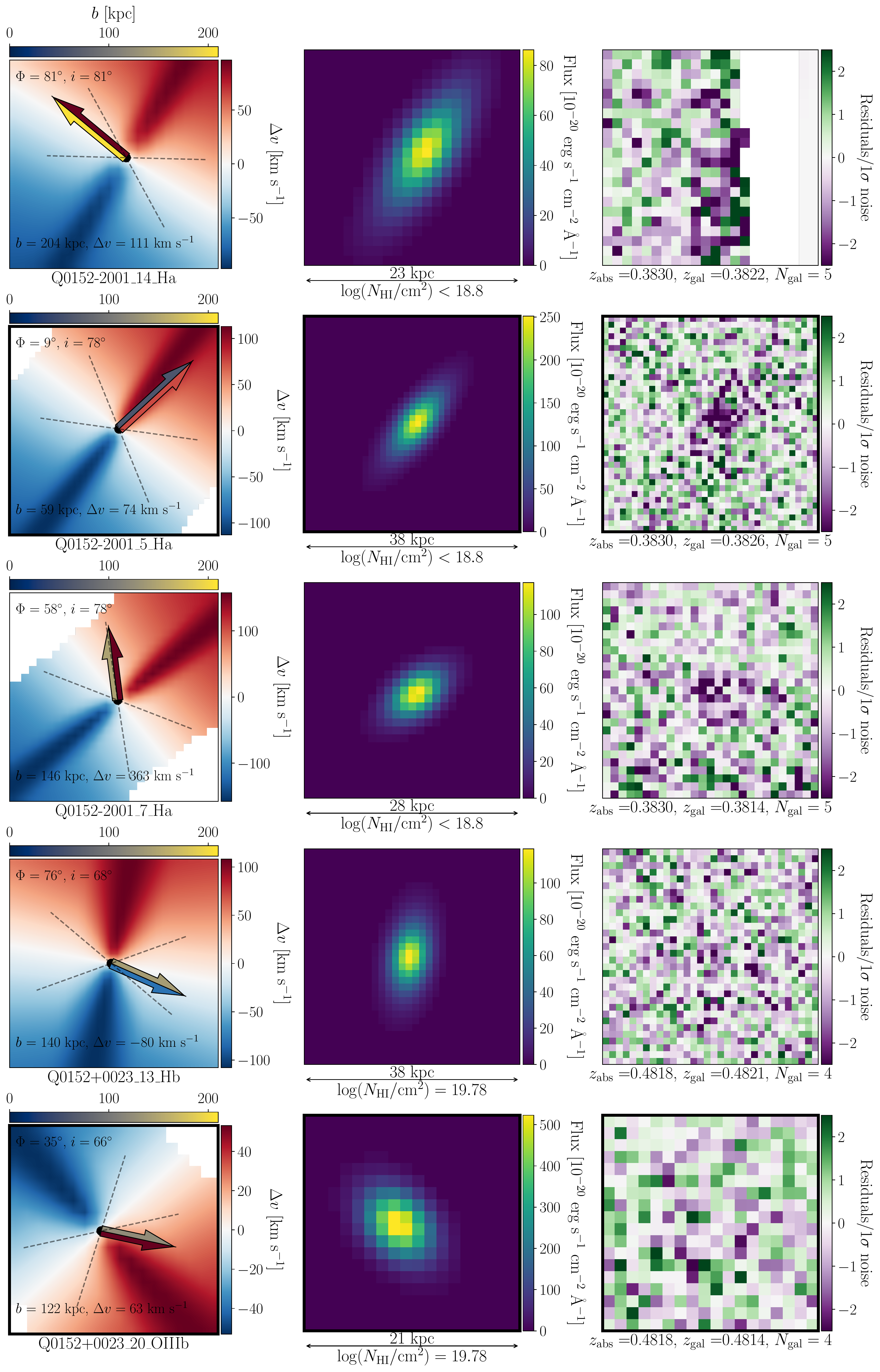}
    \caption{Continuation of \autoref{fig:Maps1}, refer to its caption. 
    }
    \label{fig:Maps2}
\end{figure*}

\begin{figure*}
    \includegraphics[width=0.81\textwidth]{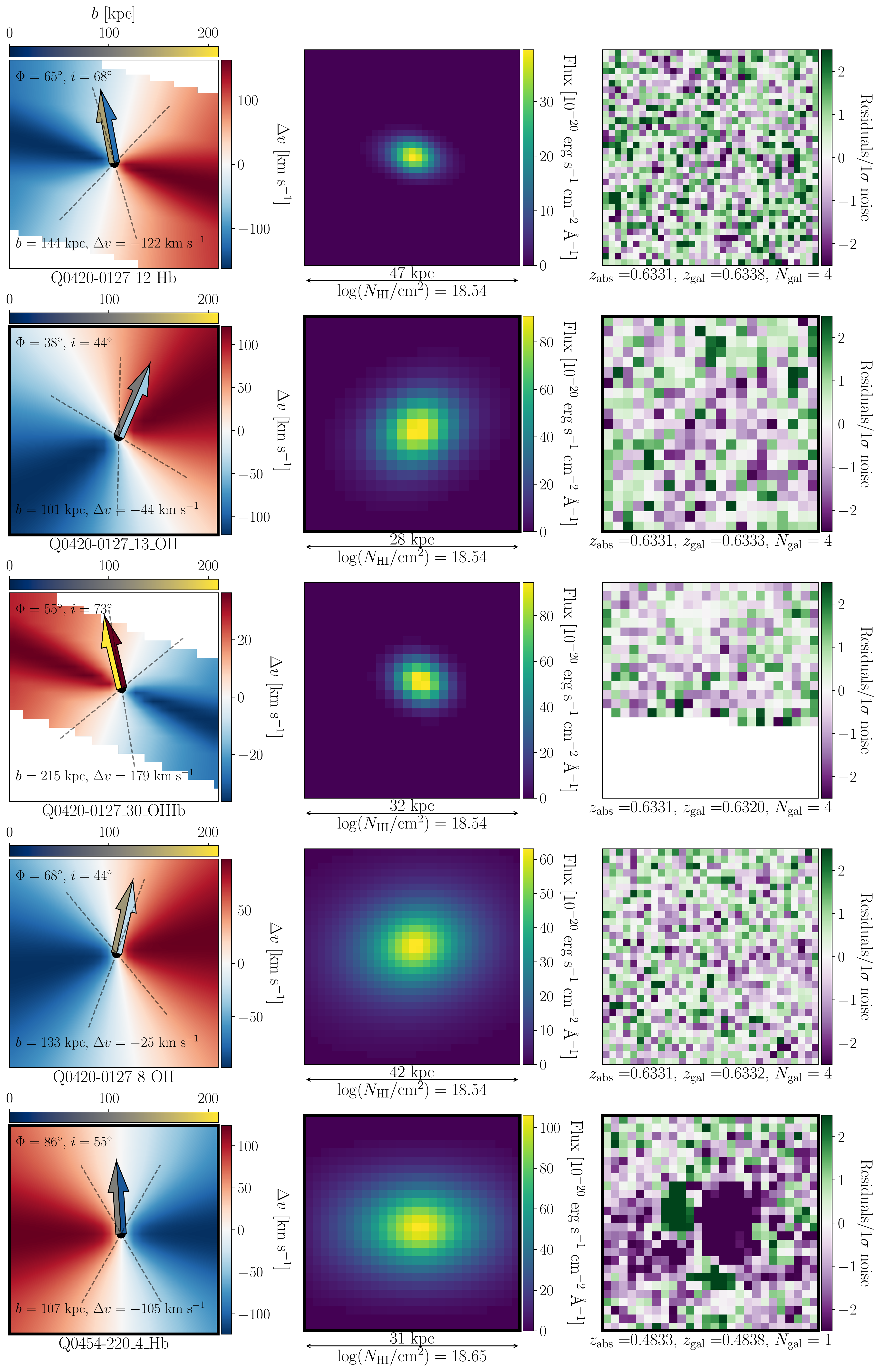}
    \caption{Continuation of \autoref{fig:Maps1}, refer to its caption. 
    The significant flux residuals for object Q0454$-$220\_4 shown in the last row are caused by the saturation of a nearby bright star in the MUSE field. 
    }
    \label{fig:Maps3}
\end{figure*}

\begin{figure*}
    \includegraphics[width=0.81\textwidth]{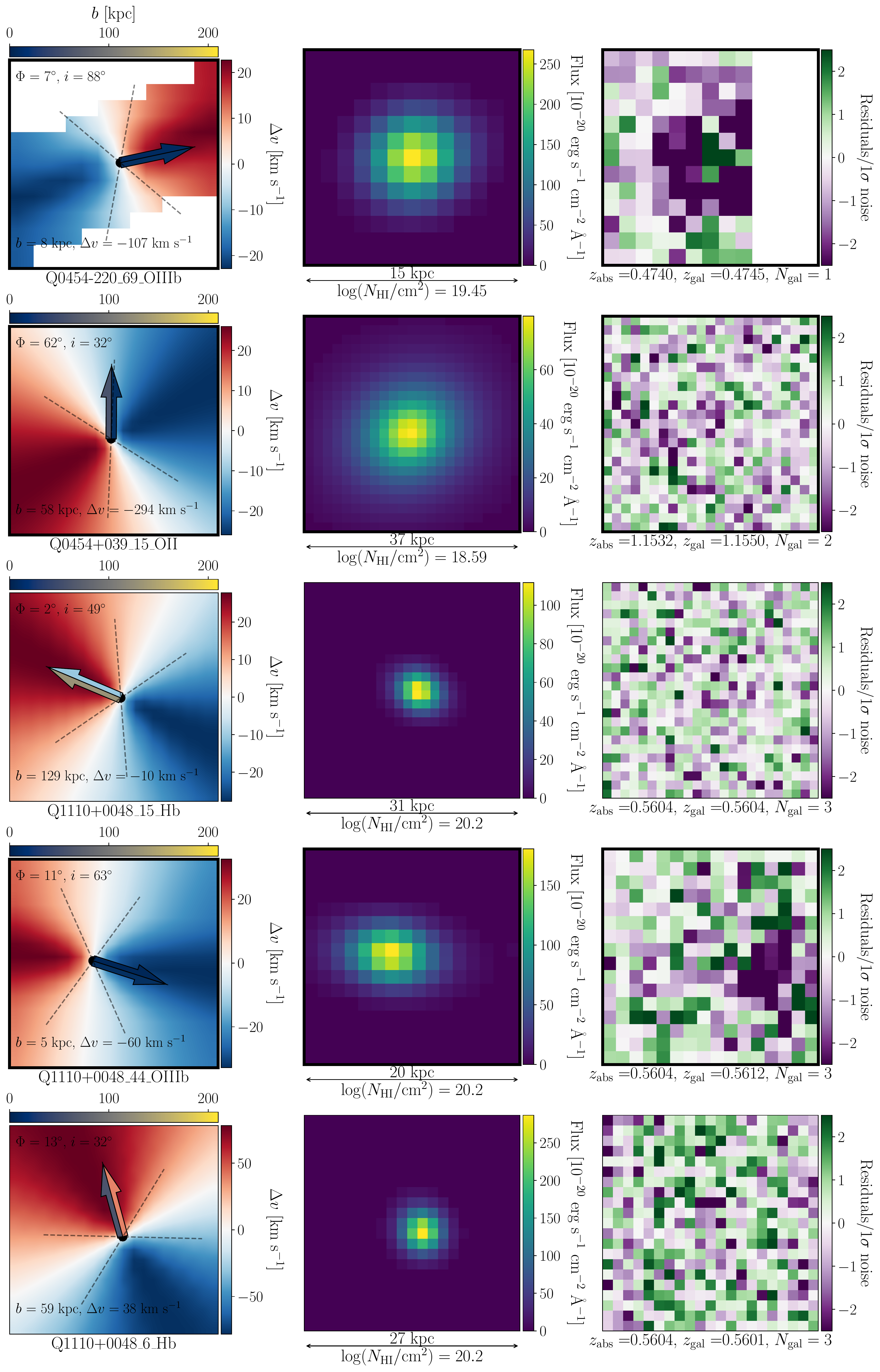}
    \caption{Continuation of \autoref{fig:Maps1}, refer to its caption. 
    Object Q0454$-$220\_69 (top row) is located less than two arcseconds from the background QSO. 
    The residuals arise from the QSO flux contaminating the galaxy flux.  
    }
    \label{fig:Maps4}
\end{figure*}

\begin{figure*}
    \includegraphics[width=0.81\textwidth]{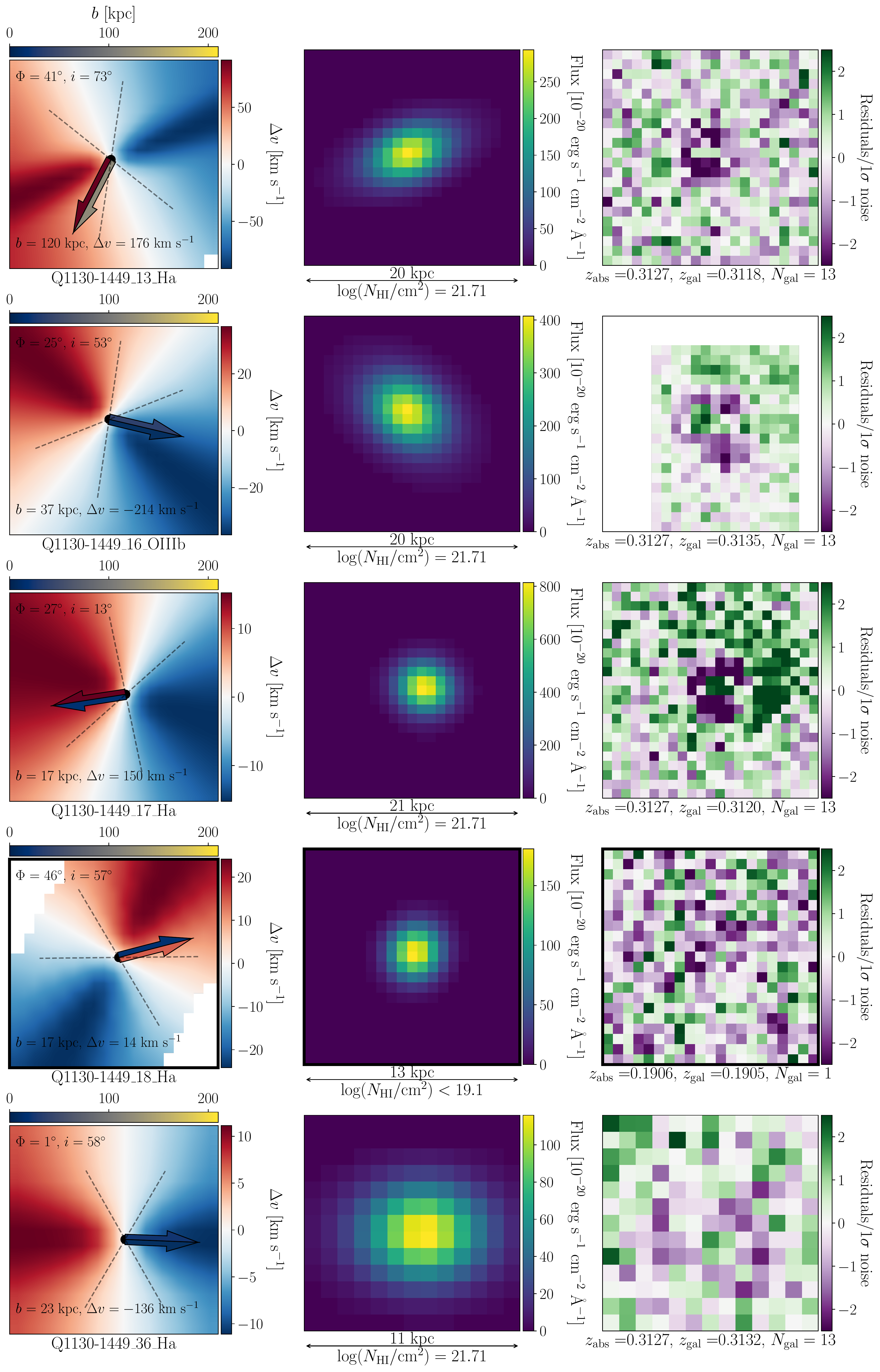}
    \caption{Continuation of \autoref{fig:Maps1}, refer to its caption. 
    The galaxies in the top three rows have significant flux residuals due to the ionizing nebula associated with a galaxy group at the absorber redshift. 
    }
    \label{fig:Maps5}
\end{figure*}

\begin{figure*}
    \includegraphics[width=0.81\textwidth]{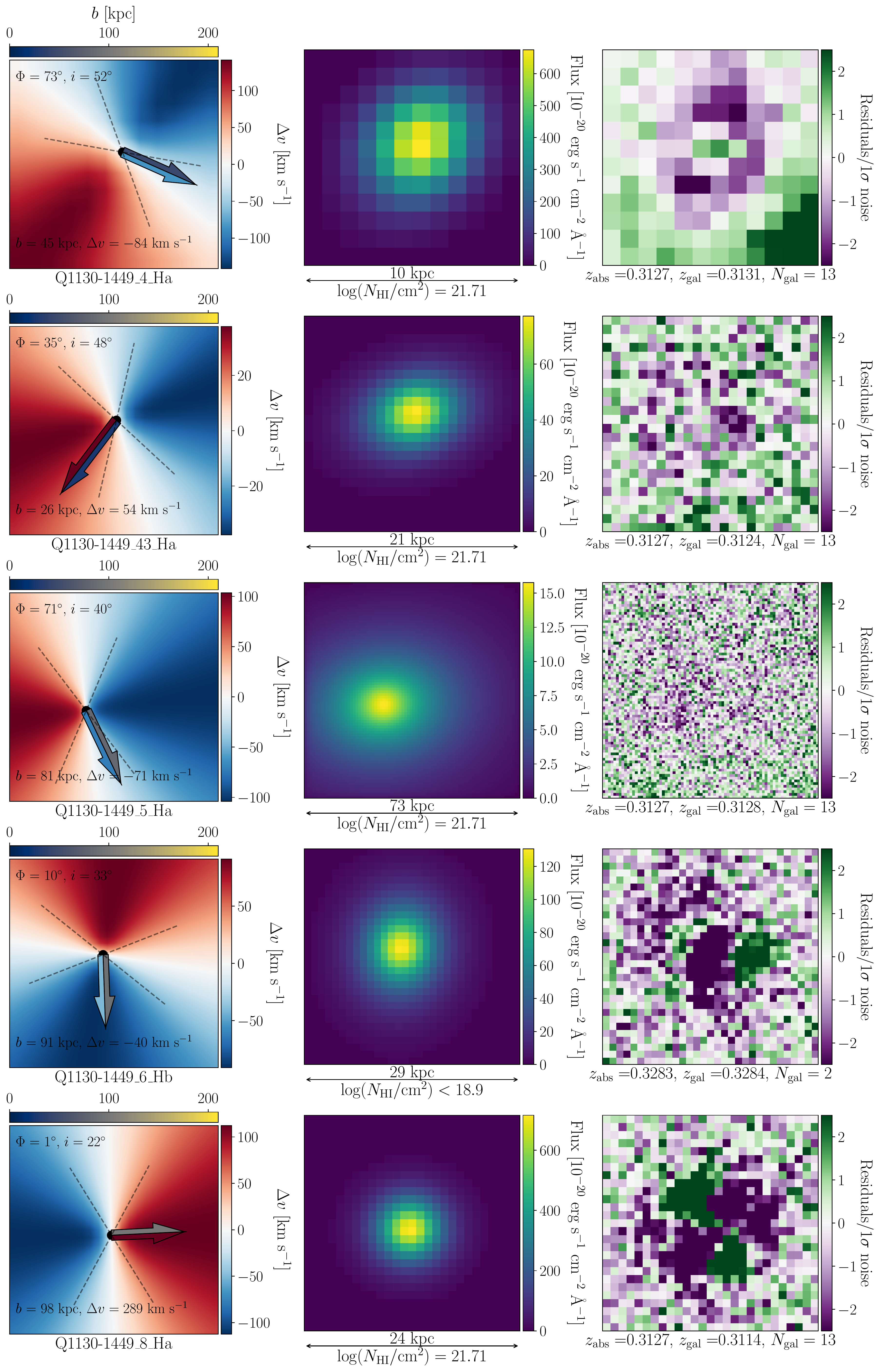}
    \caption{Continuation of \autoref{fig:Maps1}, refer to its caption. 
    The galaxy in the top row is similarly affected by the emission-line flux of the ionizing nebula. 
    In addition, galaxies Q1130$-$1449\_6 and Q1130$-$1449\_8 have asymmetric emission-line profiles and the flux residuals are likely caused by a merger and strong outflows respectively. 
    }
    \label{fig:Maps6}
\end{figure*}

\begin{figure*}
    \includegraphics[width=0.81\textwidth]{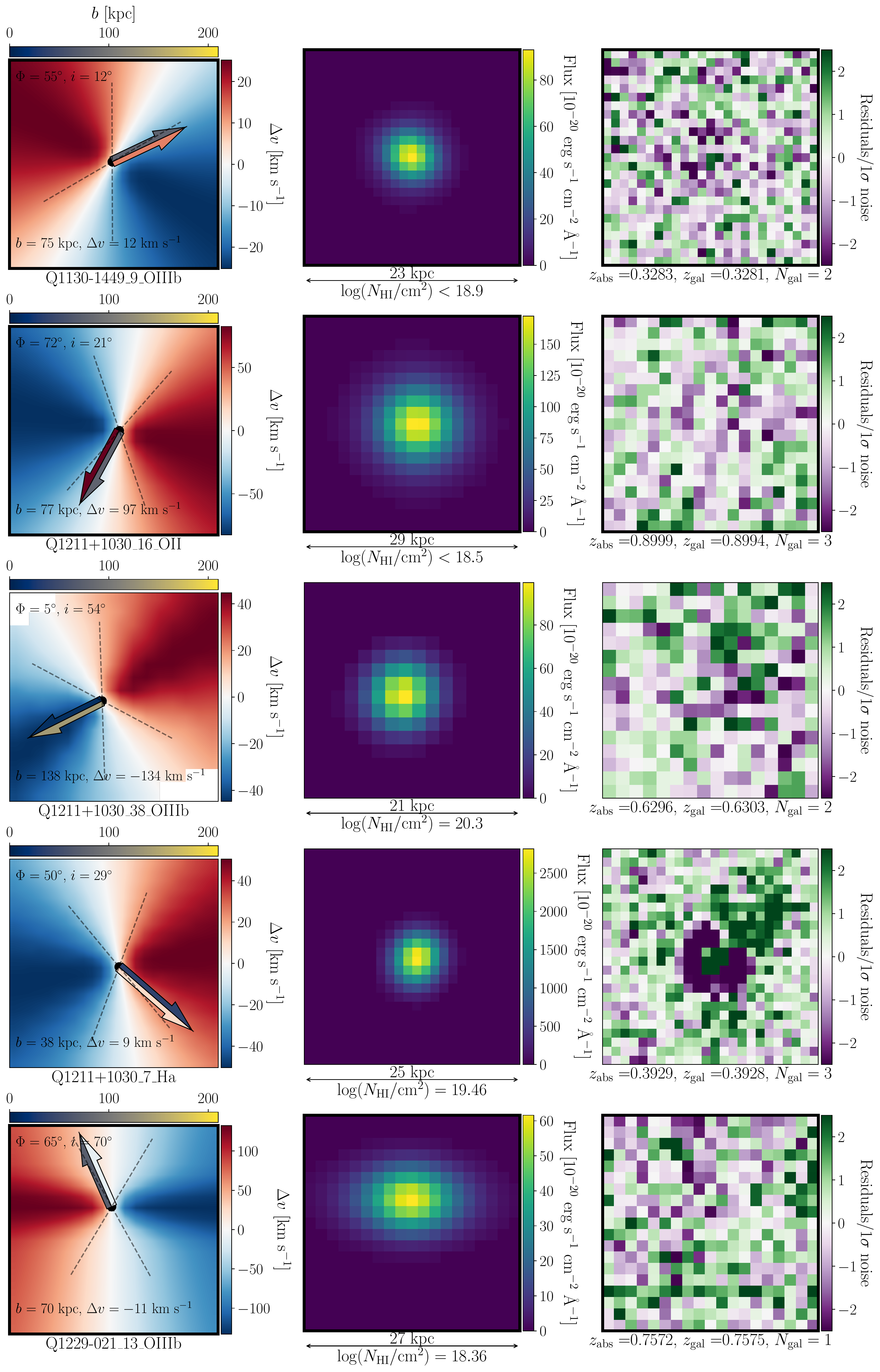}
    \caption{Continuation of \autoref{fig:Maps1}, refer to its caption. 
    Galaxy Q1211$+$1030\_7 likely hosts an active galactic nucleus from its position on the [\ion{O}{iii}]/H$\beta$ versus [\ion{O}{ii}]/H$\beta$ classification diagram \citep{Lamareille2010, Weng2022}. 
    The flux excess away from the galaxy disk is likely caused by the AGN ionizing gas. 
    }
    \label{fig:Maps7}
\end{figure*}

\begin{figure*}
    \includegraphics[width=0.81\textwidth]{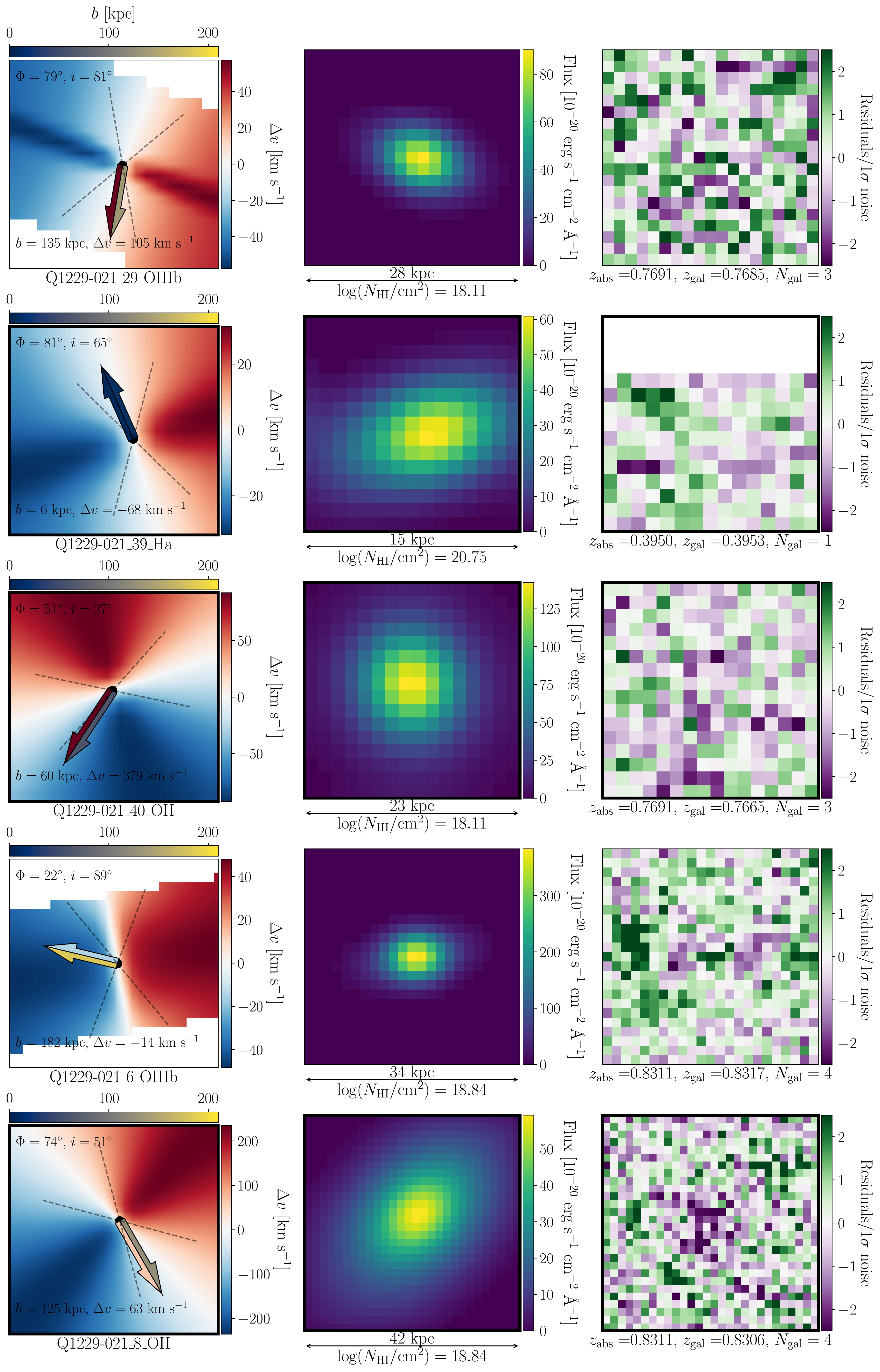}
    \caption{Continuation of \autoref{fig:Maps1}, refer to its caption. 
    The flux residuals seen in the fit for Q1229$-$021\_6 are likely also caused by an AGN \citep{Weng2022}. 
    }
    \label{fig:Maps8}
\end{figure*}

\begin{figure*}
    \includegraphics[width=0.81\textwidth]{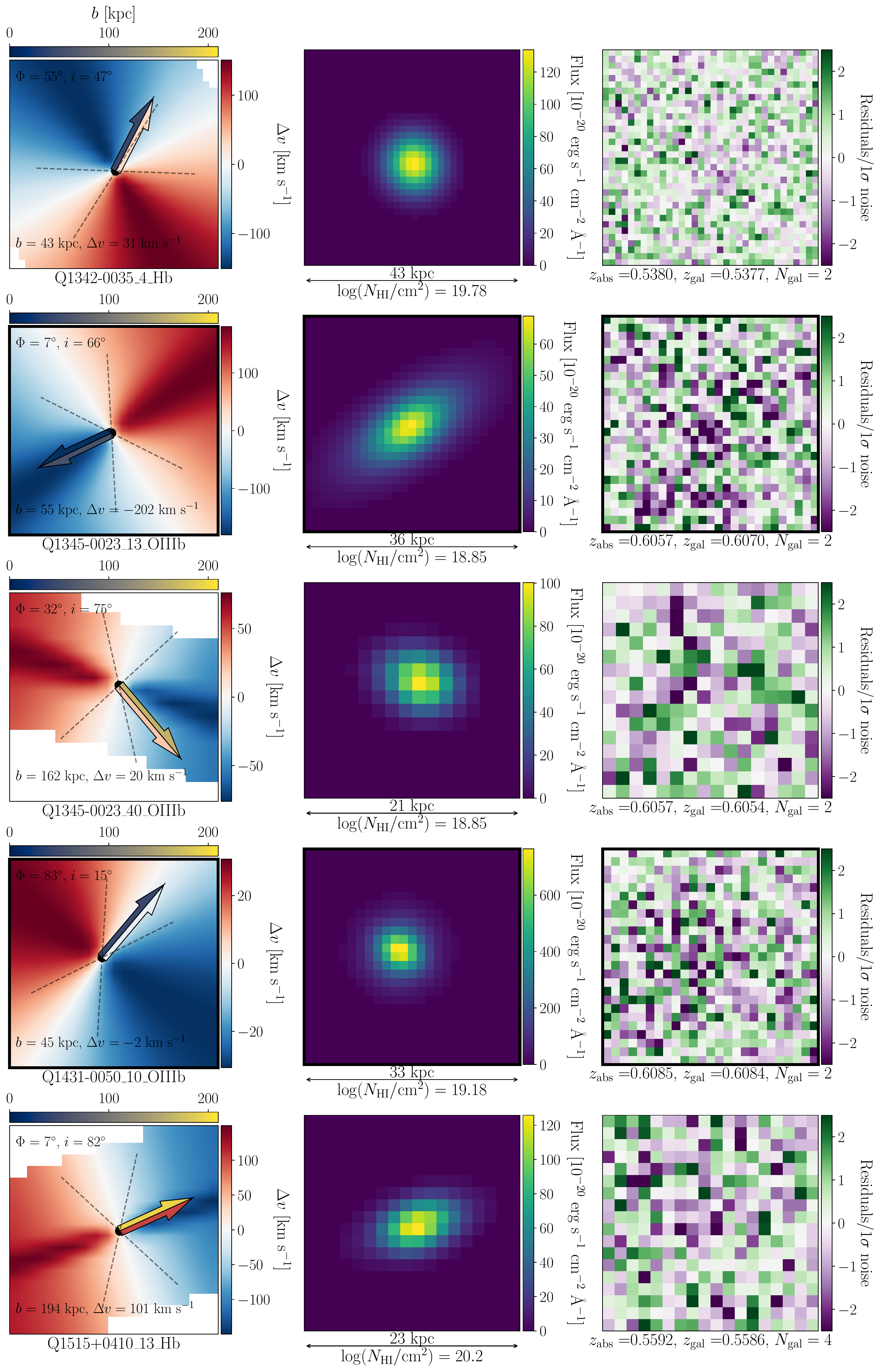}
    \caption{Continuation of \autoref{fig:Maps1}, refer to its caption. 
    }
    \label{fig:Maps9}
\end{figure*}

\begin{figure*}
    \includegraphics[width=0.81\textwidth]{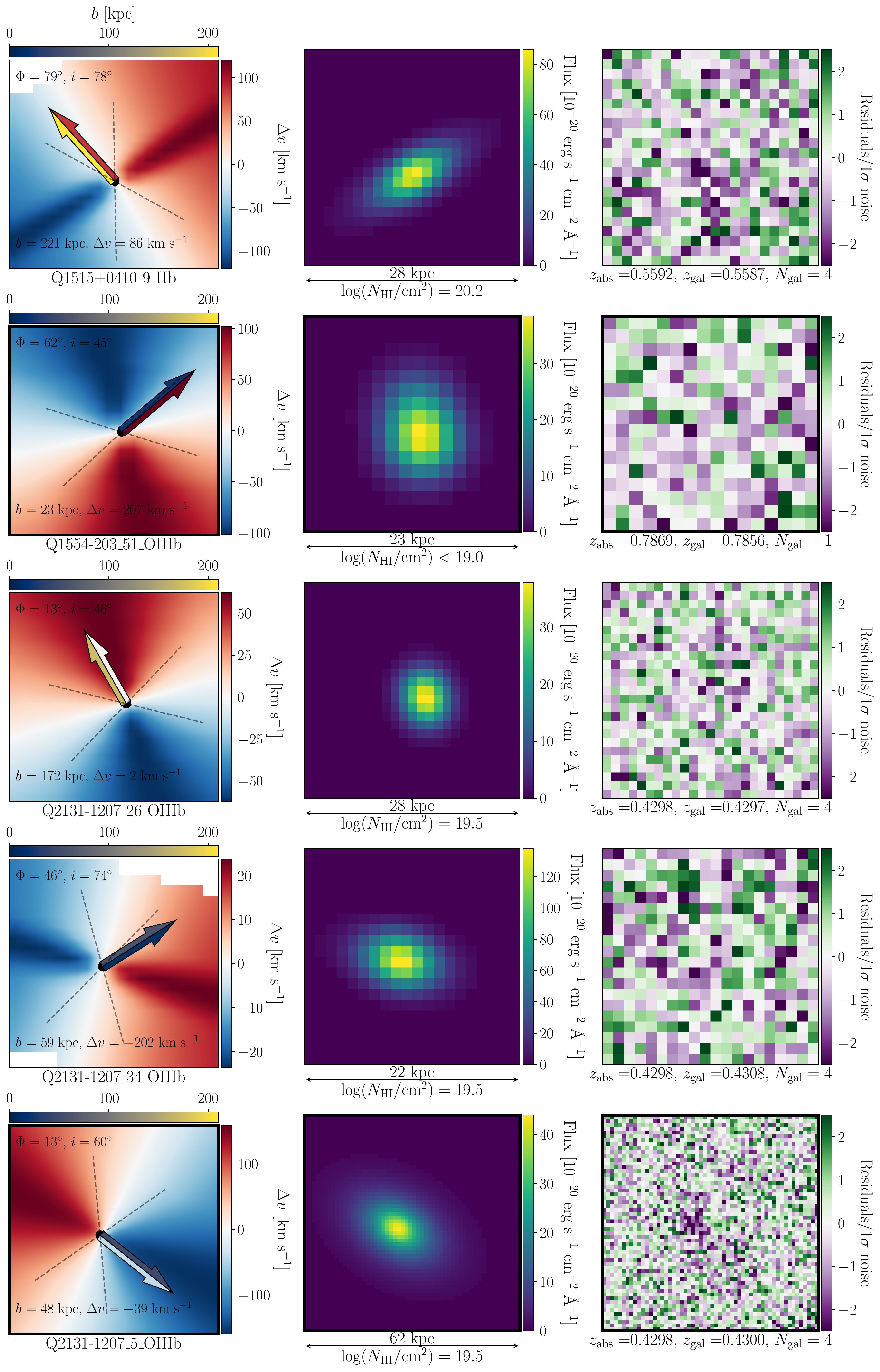}
    \caption{Continuation of \autoref{fig:Maps1}, refer to its caption. 
    }
    \label{fig:Maps10}
\end{figure*}

\section{Flow Classifications}
\label{app:flows}
In this section, we present the criteria used to classify absorbers as belonging to the galaxy disk, inflowing or outflowing. 
Additionally, we explore other origins for the gas such as the intragroup medium or low-mass, quiescent satellite at low impact parameters from the absorber that is not detected. 
We also discuss the origins of the absorber for more ambiguous cases where the data is not sufficient to draw a conclusion, multiple phenomena are possible or the absorber might trace tidally stripped gas. 
Of the 27 absorbers with associated galaxies in the MUSE-ALMA Halos survey, five have their associated galaxies modelled in previous works \citep{Rahmani2018a, Rahmani2018b, Klitsch2018, Peroux2019, Peroux2017, Szakacs2021}. 
We also note that in \citet{Hamanowicz2020}, the stellar properties of galaxies associated with a further nine absorbers (towards Q1130$-$1449, Q1211$+$1030 and Q1229$-$021) were studied. 
In this section, we present the kinematic analysis of 22 absorbers not yet analysed in detail. 

\subsection{Absorbing gas associated with the galaxy disk}
\subsubsection{Q0454$+$039 $z_{\rm abs} = 0.8596$; $\logNHIunit = 20.69$}
This damped Ly-$\alpha$ absorber is associated with a single galaxy (Q0454$+$039\_57) at an impact parameter of 18 kpc. 
The galaxy is detected in emission but is too faint to model with \textsc{galpak}. 
Given the high column density of the absorber and low impact parameter of $b < 20$ kpc, it is likely that the absorber traces gas in the galaxy disk. 
However, no kinematic nor photometric modelling is available to confirm this hypothesis. 

\subsubsection{Q1110$+$0048 $z_{\rm abs} = 0.5604$; $\logNHIunit = 20.20$}
While there are three galaxies associated with this absorber, the intervening gas is likely hosted by the galaxy at an impact parameter of $\sim$5 kpc. 
The absorber velocity aligns in sign and magnitude with the rotational velocity of the ionized gas disk (row 4 in \autoref{fig:Maps4}).  

\subsubsection{Q1211$+$1030 $z_{\rm abs} = 0.6296$; $\logNHIunit = 20.30$}
Q1211$+$1030\_57 is located at $b < 20$ kpc from the QSO and likely hosts the damped Ly-$\alpha$ absorber. 
However, there is no photometric or kinematic modelling available for this galaxy. 
Another object (Q1211$+$1030\_38) is found at $b = 140$ kpc and the absorber aligns with this galaxy's ionized gas velocity field (see row 3 in \autoref{fig:Maps7}). 
Given the characteristic radius of DLAs are $\sim$20 kpc, it is far more likely the absorber is associated with the galaxy disk of object 57. 

\subsubsection{Q1515$+$0410 $z_{\rm abs} = 0.5592$; $\logNHIunit = 20.20$}
Four galaxies are found at the redshift of this absorber, but object 4 is is at an impact parameter of $10$ kpc whereas the rest are $> 150$ kpc away. 
No kinematic modelling is available for this galaxy, but the photometric azimuthal angle is measured to be $20^\circ$. 
It is most likely that the absorber is probing dense gas in the galaxy disk. 

\subsection{Absorbing gas associated with accretion}
\subsubsection{Q1345$-$0035 $z_{\rm abs} = 0.6057$; $\logNHIunit = 18.85$}
This absorber is consistent with gas that is co-rotating with the galaxy at lowest impact parameter ($b = 56$ kpc; see the second row in \autoref{fig:Maps9}). 
With an azimuthal angle of $\Phi = 7^\circ$, the alignment with the major axis suggests the gas is inflowing onto the galaxy. 

\subsection{Absorbing gas associated with outflows}
\textbf{We explore the three most probable cases of outflows first.}

\subsubsection{Q0454$+$039 $z_{\rm abs} = 1.1532$; $\logNHIunit = 18.59$}
The galaxy at lowest impact parameter to the absorber (Q0454$+$039\_15, $b = 60$ kpc) has a projected minor axis aligned with the QSO sightline ($\Phi = 62^\circ$; see the second row in \autoref{fig:Maps4}). 
Q0454$+$039\_65 is another associated galaxy that is located 130 kpc from the absorber for which we do not have kinematic modelling. 
We find that the absorber is likely tracing expelled gas from Q0454$+$039\_15. 
While the velocity of the absorber is the same sign as the galaxy velocity field, the $\Delta v_{\rm LOS} \approx -290$ \kms\! is multiple times the measured maximum rotational velocity.

\subsubsection{Q1229$-$021 $z_{\rm abs} = 0.3950$; $\logNHIunit = 20.75$}
There is a single galaxy associated with this absorber at an impact parameter of $b < 7$ kpc. 
The azimuthal angle of $\Phi = 81^\circ$ suggests the absorber traces outflowing gas with $|v_{\rm LOS}| = 70$ \kms\!\! (see the second row in \autoref{fig:Maps8}).  
Given the proximity of the sightline to the galaxy centre, the strong absorber could also trace gas in the galaxy disk. 

\subsubsection{Q1554$-$203 $z_{\rm abs} = 0.7869$; $\logNHIunit < 19.00$}
A single galaxy is found at the redshift of this absorber at an impact parameter of $b = 23$ kpc. 
The kinematic modelling reveals the absorber is found at an azimuthal angle of $\phi = 62^\circ$ and relative velocity of $\sim$200 \kms from the systemic redshift of the galaxy. 
The absorber being traced appears to be outflowing. \\

\noindent \textbf{The remaining absorbers in this section may arise from several origins.}

\subsubsection{Q0138$-$0005; $z_{\rm abs} = 0.7821$; $\logNHIunit = 19.81$}
A single galaxy was detected in the Q0138$-$0005 field within $\pm 500$ \kms of the absorber. 
We find that the QSO line of sight aligns closely with the major axis of the associated galaxy ($\Phi = 79^\circ$; see the first row in \autoref{fig:Maps1}). 
The zinc abundance is measured to be [Zn/H]$= 0.28 \pm 0.16$, perhaps indicating a metal-enriched outflow. 
However, the line of sight velocity difference between absorber and galaxy is only $\sim$40 \kms\!. 
The true velocity of the absorbing gas may be much larger as the disk has an inclination of $i = 66^\circ$, but the seeing conditions for this observation ($R_{\rm PSF} = 1.11$ arcseconds) mean we cannot rule out a faint quiescent galaxy nearer the strong \ion{H}{i} absorber. 

\subsubsection{Q1229$-$021 $z_{\rm abs} = 0.7572$; $\logNHIunit = 18.36$}
A single galaxy is found at the absorber redshift from the MUSE data. 
The absorber appears aligned with the minor axis of object 13 in this field ($\Phi = 65^\circ$ from the last row in \autoref{fig:Maps7}). 
However, the absorber velocity is within $15$ \kms of the systemic redshift of its galaxy counterpart and may be static gas in the halo.  
It is also possible that the velocity component orthogonal to the sightline is large because of the galaxy is nearly viewed edge-on ($i = 70^\circ$). 

\subsubsection{Q1229$-$021 $z_{\rm abs} = 0.7691$; $\logNHIunit = 18.11$}
There are three galaxies associated with this absorber with impact parameters ranging from 60 to 180 kpc. 
Curiously, the absorber has the opposite velocity sign to the ionized gas velocity for object Q1229$-$021\_40 at lowest impact parameter. 
This velocity difference approaches $400$ \kms and there is no clear explanation for how this absorber could arise if it is tied to this galaxy. 
Such large velocity differences can often equate to physical distances greater than one Mpc \citep{Ho2021}. 
A possible origin for the absorber is that is traces gas being expelled from galaxy Q1229$-$021\_29 ($\Phi = 79^\circ$, see the middle row in \autoref{fig:Maps8}), but we also may be tracing gas in the intragroup medium.  

\subsubsection{Q1229$-$021 $z_{\rm abs} = 0.8311$; $\logNHIunit = 18.84$}
There are four associated galaxies located at impact parameters ranging from 125 to 187 kpc. 
The orientation and velocity of the absorber is consistent with outflowing gas from the nearest galaxy, Q1229$-$021\_8 (see the last row in \autoref{fig:Maps8}). 
An azimuthal angle of $\Phi = 74^\circ$ and velocity difference of $\Delta v_{\rm LOS} = 60$ \kms are measured from the kinematic modelling. 
However, we also cannot rule out the absorber tracing gas in the intragroup medium.



\subsection{Absorbing gas of unclear origin}
\subsubsection{Q0152$+$0023 $z_{\rm abs} = 0.4818$; $\logNHIunit = 19.78$}
We find four galaxies associated with this absorber, with impact parameters ranging from 120 to 190 kpc. 
For the nearest galaxy, Q0152$+$0023\_20, to the QSO sightline, the velocity of the absorber is consistent with co-rotating gas. 
The absorber also aligns with the ionized gas map of another galaxy (Q0152$+$0023\_13) $\sim$10 kpc away from Q0152$+$0023\_20. 
This can be seen in the last row of \autoref{fig:Maps2}. 
However, despite the consistency in velocities, the high column density absorber is outside the virial radius of both galaxies and hence, likely traces a quiescent, low-mass galaxy near the sightline. 

\subsubsection{Q0454$-$220 $z_{\rm abs} = 0.4833$; $\logNHIunit = 18.65$}
The \textsc{galpak} fits for the single galaxy detected at the absorber redshift are unreliable due to a bright nearby star in the MUSE data saturating the data (see the last row in \autoref{fig:Maps3}). 
This absorber-galaxy pair has been studied already in the literature and three more associated galaxies are located outside the MUSE field of view \citep{Kacprzak2010, Norris2021}. 
The system consists of multiple components with varying metallicities and \citet{Norris2021} finds that the gas arises from a mix of accretion, outflows and intergroup material. 
The kinematic modelling from this work does not shed further light onto the origin of the various gas components. 

\subsubsection{Q0454$-$220 $z_{\rm abs} = 0.4744$; $\logNHIunit = 19.45$}
The associated galaxy Q0454$-$220\_69 has an impact parameter of $< 10$ kpc and we expect the absorber to be tracing gas in the galaxy disk. 
However, the absorber appears to be counter-rotating with a velocity of $\sim$100 \kms from the systemic redshift of the galaxy (see first row in \autoref{fig:Maps4}). 
The modelled fit is unreliable because of the small galaxy size and nearby QSO contaminating the fit. 
A manual inspection of the [\ion{O}{iii}] emission line in the MUSE cube reveals that the galaxy emission flux is orientated such that the side with positive velocity is towards the west. 
Hence, the absorber is indeed counter-rotating, but the derived kinematic properties from \textsc{galpak} are not reliable. 
There is also no detection in the HST broadband imaging available for this continuum-faint object. 

\subsubsection{Q1130$-$1449 $z_{\rm abs} = 0.1906$; $\logNHIunit < 19.10$}
There is a single galaxy (Q1130$-$1449\_18) detected at the redshift of this absorber. 
The velocity of the ionized and neutral gas phases are consistent (see the fourth row in \autoref{fig:Maps5}). 
However, the low impact parameter $b = 18$ kpc means we cannot separate gas rotating with the galaxy disk from accreting gas. 

\subsubsection{Q1130$-$1449 $z_{\rm abs} = 0.3283$; $\logNHIunit < 18.90$}
The galaxy Q1130$-$1449\_9 is found nearest the absorber at $b = 75$ kpc. 
The absorber velocity has the opposite sign to the ionized gas velocity and is found near the projected minor axis ($\Phi = 55^\circ$, first row in \autoref{fig:Maps7}). 
However, the measured inclination of $i = 12^\circ$ suggests the galaxy is almost face-on and it becomes difficult to assign the absorber with any particular process. 
Instead, we may be tracing gas in the intragroup medium as there is another galaxy at an impact parameter of 90 kpc. 

\subsubsection{Q1211$+$1030 $z_{\rm abs} = 0.3929$; $\logNHIunit = 19.46$}
There are three galaxies almost equidistant from this absorber on the sky (37 to 39 kpc). 
The galaxy Q1211$+$1030\_7 with available kinematic modelling likely hosts an AGN, leaving significant residuals in the modelled fit (second last row in \autoref{fig:Maps7}). 
From the photometric modelling, an azimuthal angle of $\Phi = 78^\circ$ is found for object Q1211p1030\_9. 
It is difficult to assess the origin of the absorber and an alternative hypothesis that all three galaxies is that the gas is intragroup material, especially as the absorber is almost equidistant from each galaxy. 
Objects 7 and 9 in this field are only several kiloparsecs apart, meaning tidal interactions are taking place and the stripped gas can be probed in absorption. 

\subsubsection{Q1211$+$1030 $z_{\rm abs} = 0.8999$; $\logNHIunit < 18.50$}
There are two galaxies associated with this absorber with impact parameters of 77 and 175 kpc. 
The absorber aligns with the minor axis of the galaxy at smaller $b$ ($\Phi = 72^\circ$, second row in \autoref{fig:Maps7}), but the inclination angle of $i = 21^\circ$ prevents us from making a certain judgement. 

\subsubsection{Q1342$-$0035 $z_{\rm abs} = 0.5380$; $\logNHIunit = 19.78$}
Of the two galaxies associated with this absorber, the nearest galaxy at an impact parameter of 24.3 kpc has not been modelled with \textsc{galpak}. 
From the \textsc{galfit} models of the galaxy using HST broadband imaging, an azimuthal angle of $\Phi_{\rm phot} = 58^\circ$ is measured. 
Without the velocity field of the ionized gas, it is not possible to constrain the origin of the absorber. 
From the velocity map of object 4 at an impact parameter of 44 kpc, the absorber velocity is opposite in sign to the modelled ionized gas velocity. 

\subsubsection{Q1431$-$0050 $z_{\rm abs} = 0.6868$; $\logNHIunit = 18.40$}
There are two galaxies at this absorber's redshift. 
Both do not have kinematic modelling and from the photometric fitting, it is not possible to make a convincing case for the origin of the gas. 

\subsubsection{Q1431$-$0050 $z_{\rm abs} = 0.6085$; $\logNHIunit = 19.18$}
From the second last row of \autoref{fig:Maps5}, we see that the absorber aligns with the minor axis of the galaxy at closest impact parameter ($b = 45$ kpc). 
However, the galaxy appears almost face-on and the absorber is found at the systemic velocity of the galaxy. 
Hence, the origin of the gas is unclear. 



\bsp	
\label{lastpage}
\end{document}